\documentclass[twocolumn]{aastex631}
\usepackage{graphics}
\usepackage{color}
\usepackage{topcapt}
\usepackage{amsmath,amstext}
\usepackage{empheq} 
\newcommand{\etal}{et\,al.}
\newcommand{\ltsimeq}{\la}
\newcommand{\gtsimeq}{\ga}

\newcommand{\msun}{M$_{\odot}$}
\newcommand{\mstar}{$M_*$}

\newcommand{\tninety}{$\tau_{90}$}

\newcommand{\hi}{H{\sc i}}

\newcommand{\peg}{Pegasus~W}

\definecolor{amaranth}{rgb}{0.9, 0.17, 0.31}

\shortauthors{McQuinn et al.}
\shorttitle{\peg}

\begin{document}
\title{\peg: An Ultrafaint Dwarf Galaxy Outside the Halo of M31 Not Quenched by Reionization}

\author[0000-0001-5538-2614]{Kristen.~B.~W. McQuinn}
\affiliation{Department of Physics and Astronomy, Rutgers, The State University of New Jersey, 136 Frelinghuysen Rd, Piscataway, NJ 08854, USA}
\email{kristen.mcquinn@rutgers.edu}

\author[0000-0002-1200-0820]{Yao-Yuan Mao}
\affiliation{Department of Physics and Astronomy, Rutgers, The State University of New Jersey, 136 Frelinghuysen Rd, Piscataway, NJ 08854, USA}
\affiliation{Department of Physics and Astronomy, University of Utah, 115 South 1400 East, Salt Lake City, UT 84112, USA}

\author[0000-0003-1109-3460]{Matthew~R. Buckley}
\affiliation{Department of Physics and Astronomy, Rutgers, The State University of New Jersey, 136 Frelinghuysen Rd, Piscataway, NJ 08854, USA}

\author[0000-0003-3408-3871]{David Shih}
\affiliation{Department of Physics and Astronomy, Rutgers, The State University of New Jersey, 136 Frelinghuysen Rd, Piscataway, NJ 08854, USA}

\author[0000-0002-2970-7435]{Roger~E. Cohen}
\affiliation{Department of Physics and Astronomy, Rutgers, The State University of New Jersey, 136 Frelinghuysen Rd, Piscataway, NJ 08854, USA}

\author[0000-0001-8416-4093]{Andrew E. Dolphin}
\affiliation{Raytheon Technologies, 1151 E. Hermans Road, Tucson, AZ 85756, USA}
\affiliation{University of Arizona, Steward Observatory, 933 North Cherry Avenue, Tucson, AZ 85721, USA}

\begin{abstract}
We report the discovery of an ultrafaint dwarf (UFD) galaxy, \peg, located on the far side of the Milky Way-M31 system and outside the virial radius of M31. The distance to the galaxy is 915$^{+60}_{-91}$ kpc, measured using the luminosity of horizontal branch (HB) stars identified in Hubble Space Telescope optical imaging. The galaxy has a half-light radius ($r_h$) of 100$^{+11}_{-13}$ pc, $M_V = -7.20^{+0.17}_{-0.16}$ mag, and a present-day stellar mass of $6.5^{+1.1}_{-1.4}\times10^4$ \msun. We identify sources in the color-magnitude diagram (CMD) that may be younger than $\sim500$ Myr suggesting late-time star formation in the UFD galaxy, although further study is needed to confirm these are bona fide young stars in the galaxy. Based on fitting the CMD with stellar evolution libraries, \peg\ shows an extended star formation history (SFH). Using the \tninety\ metric (defined as the timescale by which the galaxy formed 90\% of its stellar mass), the galaxy was quenched only $7.4^{+2.2}_{-2.6}$ Gyr ago, which is similar to the quenching timescale of a number of UFD satellites of M31 but significantly more recent than the UFD satellites of the Milky Way. Such late-time quenching is inconsistent with the more rapid timescale expected by reionization and suggests that, while not currently a satellite of M31, \peg\ was nonetheless slowly quenched by environmental processes.
\end{abstract} 

\keywords{stars: color-magnitude diagrams $-$ galaxies: Local Group $-$ galaxies: dwarf} 
\section{Introduction}\label{sec:intro}
Very low-mass (\mstar\ $\ltsimeq10^5$ \msun) galaxies are expected to be numerous in the present-day universe, based on a $\Lambda$ Cold Dark Matter cosmology \citep[e.g.,][and references therein]{Kauffmann1993, Moore1999, Bullock2017}. Such low-mass systems have correspondingly low-luminosities ($M_V \ltsimeq -7.7$ mag) and are referred to as ultrafaint dwarfs \citep[UFDs; see, e.g.,][for a recent definition]{Simon2019}. The shallow potential wells of UFD galaxies make them extremely sensitive to both external perturbations \citep[such as heating from metagalactic UV background radiation and local environmental conditions; e.g.,][]{Bullock2000, Benson2002, Somerville2002, Geha2012, Brown2014, Wetzel2015, Rey2020, Akins2021, Pan2022} and internal perturbations \citep[such as stellar feedback; e.g.,][]{McQuinn2015a, McQuinn2015b, Applebaum2021}. Thus, these small galaxies make excellent laboratories in which to test physical models and galaxy formation theories, and constrain the impact of reionization on the growth of low-mass halos in the early universe. They are also critical components for tests of the hierarchical structure formation theories and dark matter \citep[e.g.,][]{Nadler2021}.

UFD galaxies went mostly undetected due to their low surface brightness, faint luminosities, and small physical sizes until $\sim\,20$ years ago. With the advent of wide-field, deeper optical surveys, such as the Sloan Digital Sky Survey \citep[SDSS;][]{York2000}, the Panoramic Survey Telescope \& Rapid Response System \citep[Pan-STARRS;][]{Chambers2016}, the Pan-Andromeda Archeological Survey \citep[PAndAS;][]{McConnachie2009}, the Dark Energy Survey \citep[DES;][]{DrlicaWagner2020}, the DECam Local Volume Exploration \citep[DELVE;][]{Cerny2022a,Cerny2022b}, and the DESI Legacy Imaging Surveys \citep{Dey2019}, searches for very low-mass galaxies have been incredibly fruitful, evidenced by the growing number of UFD galaxies now known \citep[e.g.,][and references therein]{Simon2019}. The majority of the galaxies discovered lie in close proximity to the Milky Way (MW) and the Large Magellanic Cloud (LMC), as intrinsically faint systems are more readily detected at closer distances. A smaller number have been identified as part of the M31 satellite system, found via the targeted PAndAS observations \citep{McConnachie2009} or from DESI data \citep{Collins2022}, and, more recently, a galaxy was found outside the Local Group \citep{Sand2022}.

Here, we report the discovery of an UFD galaxy in the Local Group (LG) named \peg. The galaxy was identified in the DESI Legacy Imaging Surveys data as an over-density in the photometric stellar catalog, similar to previous discoveries \citep[e.g.,][]{Bechtol2015, Drlica-Wagner2015, Koposov2015, Collins2022}, but the system is much farther than other known UFD galaxies in the LG. {\em Hubble Space Telescope (HST)} optical imaging of \peg\ enables a robust distance measurement which places the galaxy on the far side of M31 but outside M31's virial radius. Thus, \peg\ offers a unique opportunity to measure the properties and study the evolution of an UFD galaxy that is not a satellite galaxy but that is still in close enough proximity for detailed observations.

The paper is organized as follows. Section~\ref{sec:data} describes the {\em HST} observations, photometry, and data processing. Section~\ref{sec:structure} presents the structural parameters of \peg. Section~\ref{sec:cmd} explores the CMD of the galaxy. Section~\ref{sec:distance} includes the distance measurement to the galaxy based on horizontal branch (HB) stars and investigates the LG environment around \peg. Section~\ref{sec:sfh} presents the SFH based on fitting stellar evolution libraries to the CMD, calculates star formation timescales, and measures the integrated luminosity and stellar mass of the galaxy. Section~\ref{sec:discuss} summaries the overall properties of \peg\ in the context of other Local Group UFDs and discusses the implications of an UFD galaxy that was not quenched by reionization. 

\begin{figure*}
\begin{center}
\includegraphics[width=0.32\textwidth]{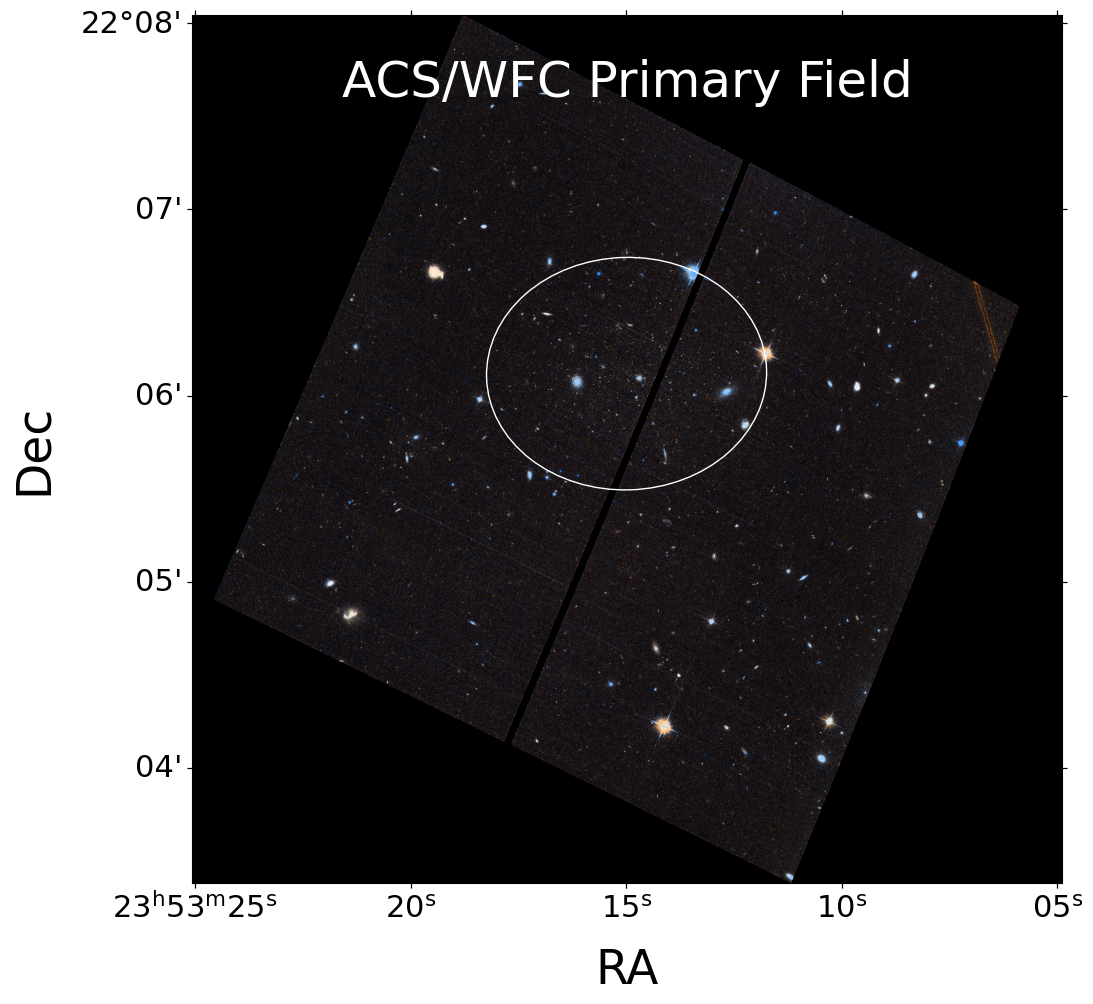}
\includegraphics[width=0.33\textwidth]{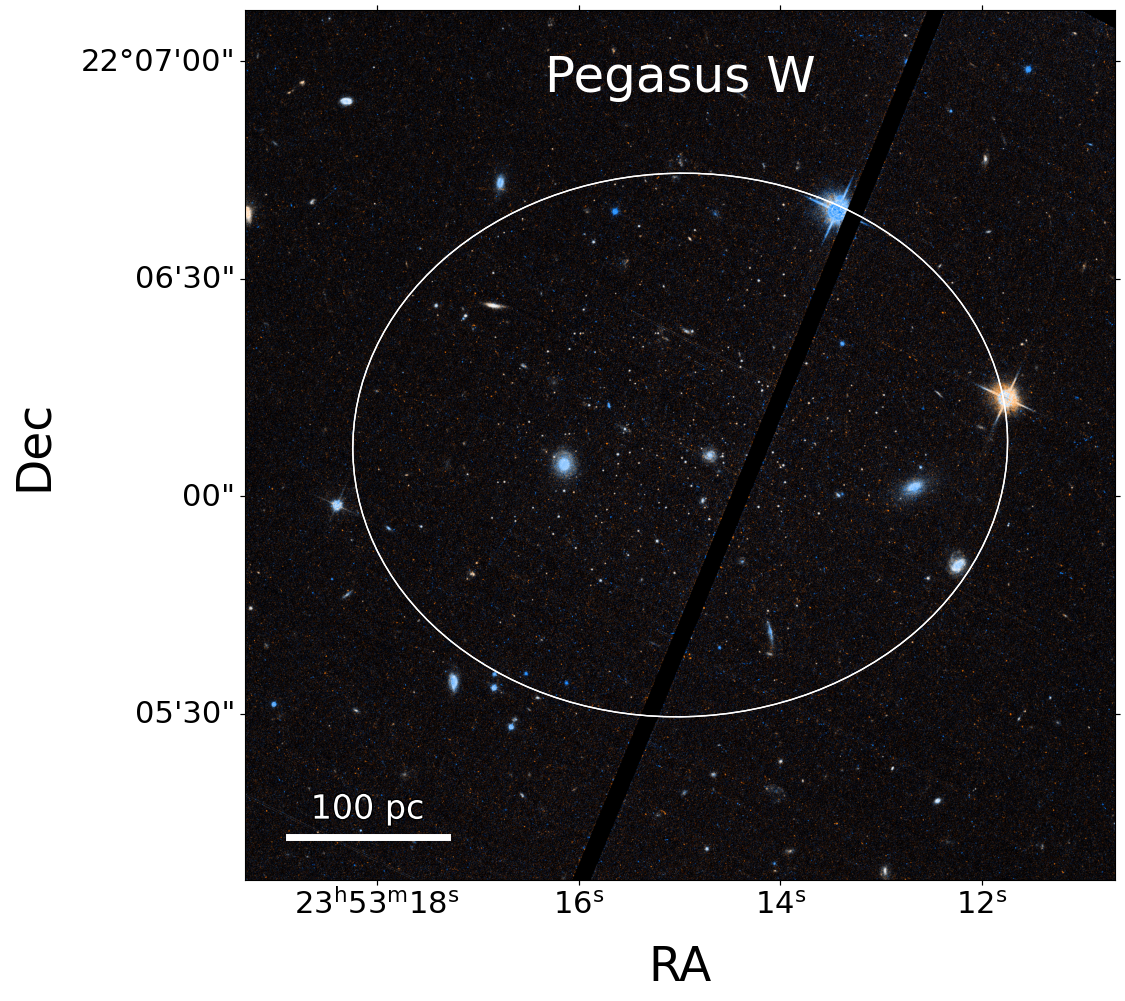}
\includegraphics[width=0.32\textwidth]{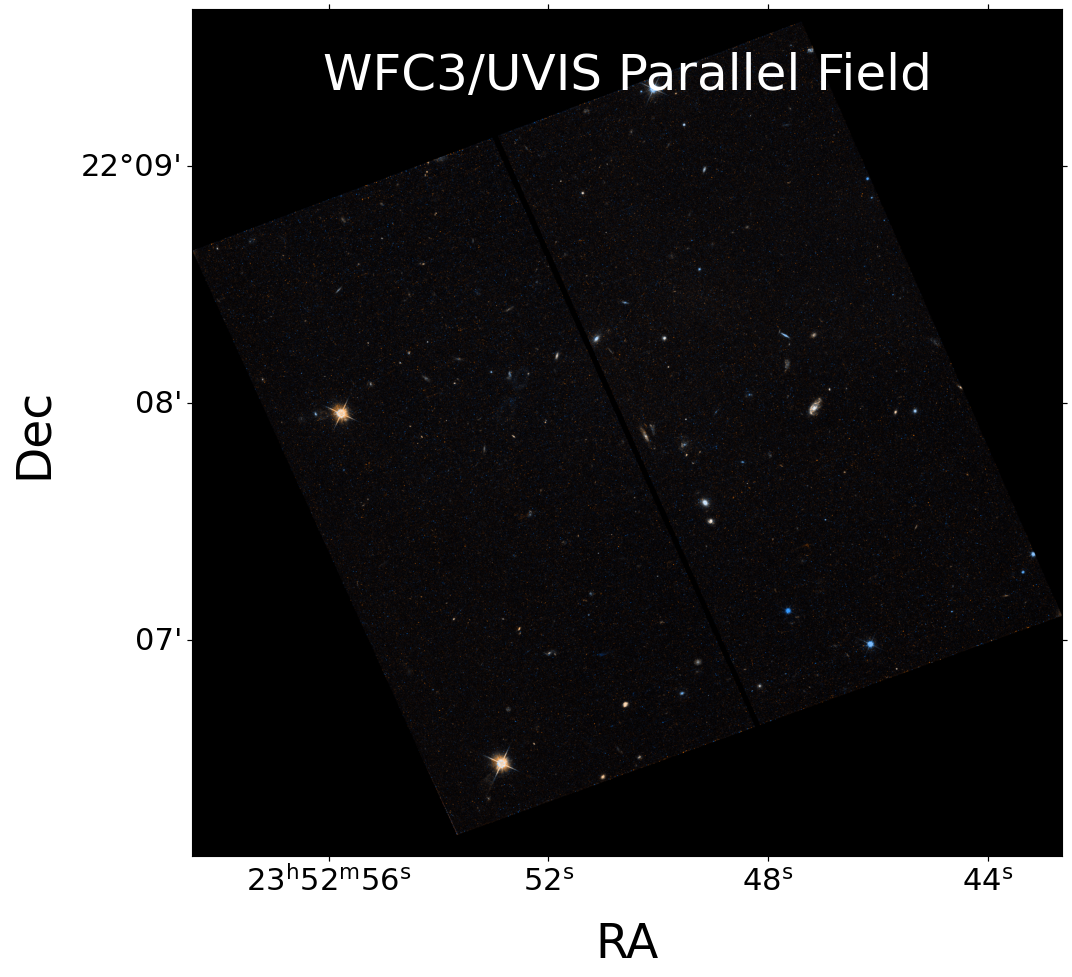}
\end{center}
\caption{Three-color images of the observations. Left: The ACS full field of view with an ellipse encircling the stellar component of \peg\ out to $2~r_h$ based on the best-fitting structural parameters. Middle: A zoom-in on the region of the galaxy. Right: The WFC3 parallel field which can help quantify potential foreground and background contamination. The three-color images were created using F606W for red, and average of F606W and F814W images for green, and F814W for blue.}
\label{fig:image}
\end{figure*}

\section{Observations and Data Processing}\label{sec:data}
\subsection{Observations}
{\em HST} observations of \peg\ were obtained using the Wide Field Camera (WFC) of the Advanced Camera for Surveys (ACS) instrument \citep{Ford1998} on 2022 June 27 in the F606W and F814W filters as part of HST-GO-16916. All the {\it HST} data used in this paper can be found in MAST: \dataset[10.17909/x8qj-bn51]{http://dx.doi.org/10.17909/x8qj-bn51}. Two observations were made per filter during one orbit, with a 5$\times$5 pixel dither ({\tt acs wfc dither line pattern \#14}) between exposures to help reject cosmic rays and to mitigate detector cosmetic defects. The ACS pointing was centered on J2000 $\text{RA} = 23$\,:\,53\,:\,14.229, $\text{Dec} =+22$\,:\,05\,:\,35.54, which placed the galaxy slightly offset from the center of the ACS field of view. This selected pointing ensured maximum coverage of the stellar disk while avoiding nearby bright stars. The exposure time was evenly split between the 
two ACS/WFC filters with a final integration time of 1140 s per filter. 

A second field was simultaneously imaged in parallel using the Wide Field Camera 3 (WFC3) UVIS instrument in the same bandpasses, enabling an investigation of potential background and foreground contamination in a field near to the galaxy. The WFC3 pointing was centered on $\text{RA}=23$\,:\,52\,:\,51.255, $\text{Dec}=+22$\,:\,07\,:\,48.48 with integration times of 1020 and 1045 s in F606W and F814W filters, respectively.

Figure~\ref{fig:image} presents 3-color images of the {\em HST} ACS observations (left), a zoom-in on the region with \peg\ (center), and the parallel field from the WFC3 observations (right). The 3-color images were made using the F606W, F814W, and (F606W+F814W)/2 mosaics created from the charge transfer efficiency corrected ({\tt flc.fits}) files with the {\em HST} {\tt drizzlepac v3.0} python package \citep{Hack2013, Avila2015}. Despite being low-luminosity, the galaxy is visible in the center image as an over-density of point sources. 

\begin{table}
\begin{center}
\caption{Pegasus W Properties}
\label{tab:properties}
\end{center}
\begin{center}
\vspace{-15pt}
\begin{tabular}{lr}
\hline 
\hline 
Property				& Value \\
\hline
RA (J2000) 				& 358.31248167$^{\circ}\pm$1\arcsec \\
Dec (J2000)				& 22.10197022$^{\circ}\pm$1\arcsec\\
Position angle $\theta$ ($^{\circ}$ E of N) & 92$\pm$3 \\
ellipticity ($\epsilon= 1 - \frac{b}{a}$) & 0.17$^{+0.07}_{-0.08}$ \\
$r_h$ (\arcsec)			& 23$\pm2$ \\
$r_h$ (pc)				& 100$^{+11}_{-13}$ \\
$M_V$ (mag)				& $-7.20^{+0.17}_{-0.16}$ \\
\mstar\ (\msun)			& $6.5^{+1.1}_{-1.5}\times10^4$\\
HB m$_{V, 0}$ (mag)			& $25.30^{+0.12}_{-0.20}$ \\
$\mu$	(mag)				& 24.81$^{+0.14}_{-0.22}$ \\
$\Sigma_b$ (arcmin$^{-2}$)		& 16.8$^{+4.1}_{-15.4}$ \\
Distance (kpc)			& 915$^{+60}_{-91}$\\
$[$M/H$]$ (dex)	 		& $-1.9\pm0.1$ \\
\tninety\ (Gyr)   	      		& $7.4^{+2.2}_{-2.6}$ \\ 
$A_V$ (mag)				& 0.315  \\
$A_{F606W}$ (mag)			& 0.284 \\
$A_{F814W}$ (mag)			& 0.175  \\
\hline
\hline              
\end{tabular}
\end{center}
\tablecomments{The properties of Peg~W were measured in this work, with the exception of the foreground extinction which is from \citet{Schlegel1998} with recalibration from \citet{Schlafly2011}. }
\end{table}

\subsection{Photometry}
Photometry was performed on the {\tt flc.fits} images using the point spread function (PSF) fitting software {\tt DOLPHOT} \citep{Dolphin2000, Dolphin2016}, with includes specific ACS/WFC and WFC3/UVIS modules. The {\tt DOLPHOT} photometric parameters were set according to the values recommended in \citet{Williams2014, Williams2021}. 

The photometric output was filtered for well-recovered stars using a combination of  quality metrics returned from {\tt DOLPHOT} that help to characterize each source. Specifically, we selected sources with an output error flag $<4$, object type $\leq2$, signal-to-noise ratio $\geq5$ in both filters, sharp$_{F606W}^2 +$ sharp$_{F814W}^2 < 0.075$, and crowd$_{F606W} + $ crowd$_{F814W} < 0.1$. The sharpness parameter is a measure of how peaked or broad a source is relative to the PSF and helps to reject cosmic rays and background galaxies, respectively. We chose to apply strict sharpness cuts to limit contamination from faint, unresolved background galaxies in the stellar catalogs. The crowding parameter measures how much brighter a source would have been if stars nearby on the sky had not been fit simultaneously. While an important quality metric to consider, strict crowding cuts are not as critical in creating a high-fidelity catalog given the spareness of the field.

Artificial stars tests were performed on both the ACS and WFC3 data to measure the observational uncertainties and completeness of the images using the same photometric software. Approximately 500k artificial stars were injected into each individual image following to the spatial distribution of all sources identified in the photometry (i.e., the pre-filtered {\tt DOLPHOT} output). 
The sources were then recovered photometrically and the same quality cuts used for the photometry were applied to the output. The field was sufficiently uncrowded that we detected no significant trends of incompleteness with distance from the center of the galaxy.

\section{Structural Parameters}\label{sec:structure}
We determined the structural parameters of \peg, including orientation on the sky (position angle, $\theta$), shape (semi-major axis, $a$; ellipticity, $\epsilon = 1 - \frac{b}{a}$), and half-light radius ($r_h$). These parameters help characterize \peg\ and provide a way to compare to the properties of \peg\ to other UFD galaxies (see Section~\ref{sec:discuss}). They are also used to apply spatial cuts to the photometry to create our final stellar catalog for \peg. 

The structural parameters were determined using an unbinned maximum likelihood approach. Specifically, we perform a Markov Chain Monte Carlo (MCMC) fit of an exponential density profile (allowing for non-zero ellipticity) to the spatial distribution of observed sources over the full ACS/WFC field of view. To avoid any impact of incompleteness while maximizing the contribution of galaxy members, we consider only sources with F606W $<$ 26.6, F814W $<$ 25.7 (1.5 mag brightward of the 50\% completeness limit in both filters, ascertained from the artificial star tests) and a color $(\textrm{F606W} - \textrm{F814W})$ $<$ 1.5. We verified that the resulting structural parameters presented below are quite robust to these cuts, such that either eliminating the color cut entirely and/or moving the magnitude cuts faintward yielded results that were consistent to within their 1$\sigma$ uncertainties.  

Operationally, we fit for six free parameters: The tangent plane coordinates $x_{0}$ and $y_{0}$, which are the location of the center relative to an arbitrary positional zeropoint\footnote{The positional zeropoint was set to a rough guess center of (RA,Dec) = (358.312732$^{\circ}$,22.102465$^{\circ}$) based on the distribution of sources recovered photometrically.}, the half-light radius $r_{h}$, the ellipticity $\epsilon = (1-b/a)$ where $b/a$ is the ratio of minor to major axis lengths, the position angle $\theta$ in degrees east of north, and $N_{\star}$, which is the total number of stars in the galaxy (within the aforementioned CMD limits). Note that $N_{\star}$ is an extrapolation from the density profile rather than the number of stars we observe within the ACS/WFC field of view, which is $N_\text{obs} = $ 314 after applying our CMD cuts.

To determine the best-fit values of the structural parameters and their uncertainties, we search for the set of parameters for which the data are most likely.  In other words, for parameters $(p_1, p_2, ... ,p_6)$ we maximize the likelihood function:
\begin{equation}
    \mathcal{L}(p_1, p_2, ... ,p_6) = \prod_{i} \ell_{i} (p_1, p_2, ... ,p_6),
    \label{eq:like_l}
\end{equation}
 where $\ell_{i} (p_1, p_2, ... ,p_6)$ is the probability of finding the $i$th datapoint given parameters $(p_1, p_2, ... ,p_6)$.  This probability is calculated following  \citet{Martin2008, Martin2016b}, assuming the target galaxy has an exponential density profile $\rho_\text{gal}(r)$ as a function of elliptical radius $r$:
 \begin{equation}
     \rho_\text{gal}(r) =  \frac{1.68^2}{2\pi r_{h}^{2}(1-\epsilon)} N_{\star} \exp(-1.68r/r_{h}),
     \label{eq:like_r}
 \end{equation}
 where the elliptical radius $r$ is related to the tangent plane coordinates $x$ and $y$ as
 \begin{equation}
     \begin{split}
     r = \biggl\{\left[ \frac{1}{1-\epsilon} ((x-x_{0})\,\textrm{cos}\,\theta - (y-y_{0})\,\textrm{sin} \,\theta)\right]^{2} + \\
     ((x-x_{0})\,\textrm{sin}\,\theta + (y-y_{0})\,\textrm{cos}\,\theta)^{2}\biggr\}^{1/2}.
    \end{split}
 \end{equation}
 
\begin{figure}
\gridline{\fig{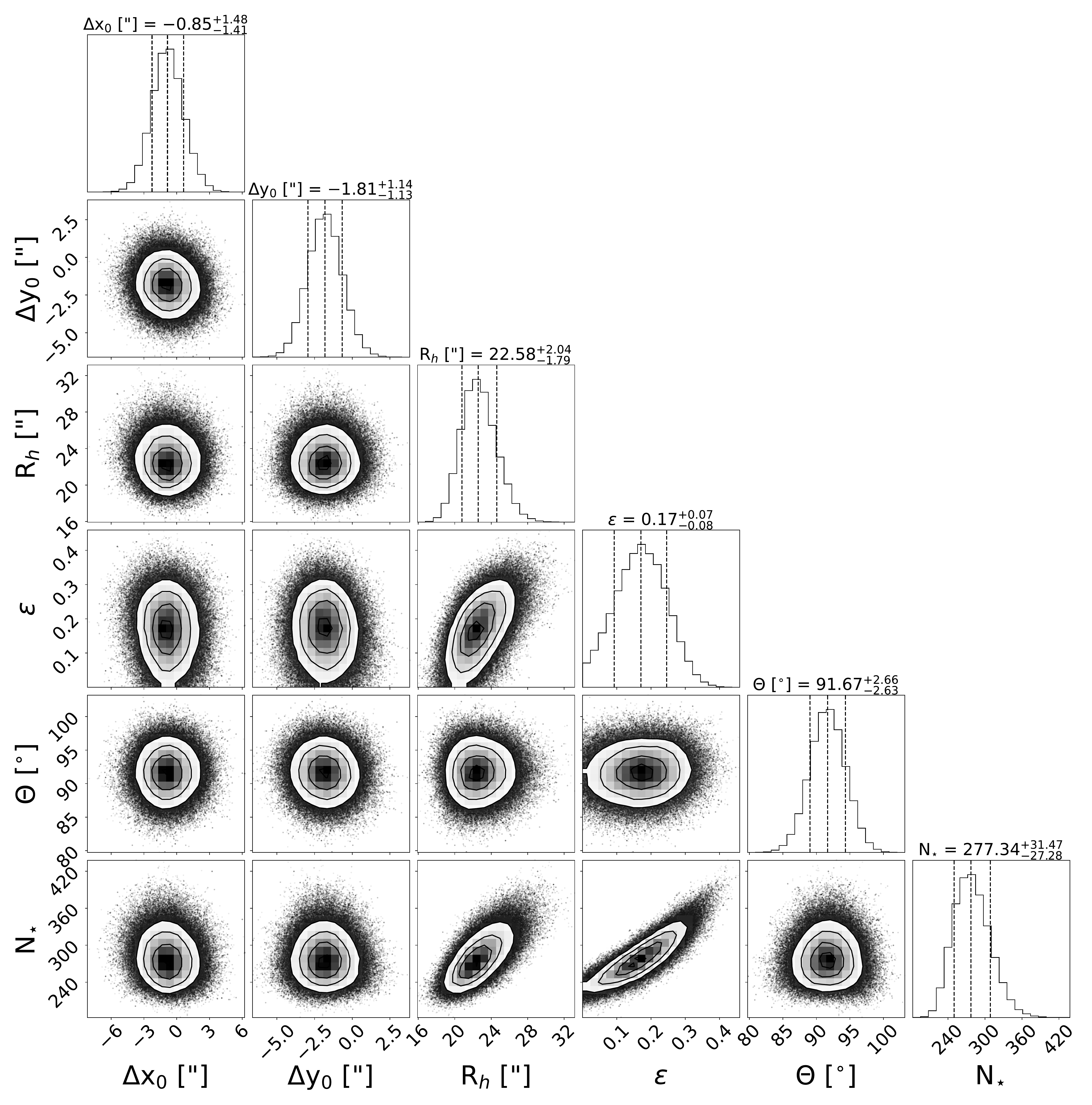}{0.49\textwidth}{}}
\vspace{-15pt}
\caption{Full posterior distributions of \peg\ structural parameters over 10,000 post-burn-in MCMC iterations.  The black contour lines correspond to 1,2 and 3$\sigma$. Best-fitting values are listed in Table~\ref{tab:properties}.}
  \label{fig:corner}
\end{figure}

We also allow for a spatially constant background density $\Sigma_{b}$.  The background density in each MCMC iteration is set by requiring that the number of background sources is equal to the difference between the model-predicted number of sources and the observed number of sources in the full field of view, with area $A$: 
\begin{equation}
    \Sigma_{b} = \left({N}_\text{obs} - \int_{A} \rho_\text{gal}\, dA\right) / A.
\end{equation}

In practice, the integration of Eq.~\ref{eq:like_r} over the field of view and the calculation of $A$ are performed numerically by generating a spatial grid of pixels with an area of 1 arcsec$^{2}$.  With $\Sigma_{b}$ in hand for each iteration, the full density profile for a given set of trial parameters is:
\begin{equation}
    \rho_\text{model}(r) = \rho_\text{gal}(r) + \Sigma_{b}.
    \label{eq:like_rhomod}
\end{equation}

Combining Eqs.~\ref{eq:like_l}-\ref{eq:like_rhomod}, the log likelihood is then:
\begin{equation}
    \textrm{ln}\,\mathcal{L} = \sum_{i}^{N_\text{obs}} \rho_\text{model}(r_{i}) - N_\text{obs}.
\end{equation}

We impose broad, flat, astrophysically motivated priors on the free parameters:
\begin{itemize}
    \item $|x_{0}|$, $|y_{0}| <$ 100$\arcsec$, essentially requiring the center of the galaxy to lie in our field of view.
    \item $r_{h} >$ 0.
    \item 0 $< \epsilon \leq$ 1.
    \item 0 $< \theta \leq \pi$.
    \item $N_{\star} > 0$.
    \item $\Sigma_{b} \geq$ 0.
\end{itemize}

We sample the posterior distributions of the parameters using the \texttt{emcee} package \citep{Foreman-Mackey2013}, running 50 affine-invariant walkers for 5000 burn-in iterations followed by 10000 production iterations (more than sufficient given autocorrelation lengths of $<$100 steps for all parameters).  The best-fit parameter values we report are the medians over all post-burn-in iterations, with uncertainties corresponding to the 16th and 84th percentiles.  

\begin{figure}
\gridline{\fig{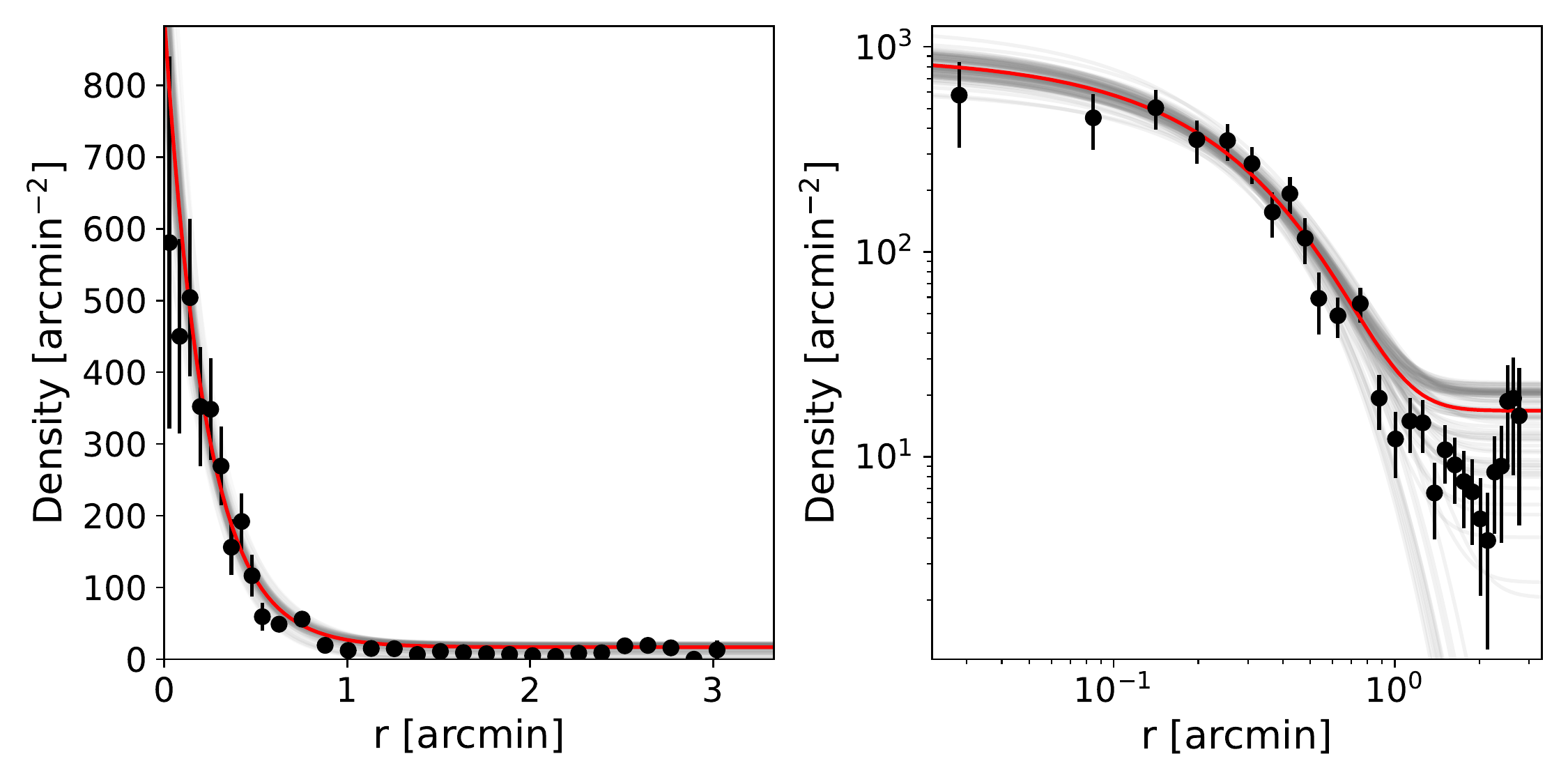}{0.47\textwidth}{}}
\vspace{-20pt}
\caption{Binned radial density profile of \peg, shown on linear (left) and logarithmic (right) scales. Black points are the binned observed data with errorbars indicating Poissonian uncertainties.  The red line is the best-fit exponential profile from our maximum likelihood analysis, and the grey lines represent 100 random draws from the posterior distributions of the profile parameters.}
  \label{fig:densprof}
\end{figure}

Figure~\ref{fig:corner} shows the full posterior distributions of our fits and their correlations.  As an additional check, we calculate a binned radial density profile in Figure~\ref{fig:densprof}, where black points are the binned observed data and vertical errorbars represent Poissonian uncertainties. Overplotted on the binned data is the maximum-likelihood solution in red, with 100 random individual draws from the posterior distributions shown in grey.

The final structural parameters including center RA, Dec coordinates of \peg, $r_h$, ellipticity, and position angle, as well as the value of $\Sigma_b$, are listed in Table~\ref{tab:properties}. We also overplot an ellipse based on these parameters and encompassing $2~r_h$ on the 3-color ACS images in Figure~\ref{fig:image}.

\section{Color-Magnitude Diagram}\label{sec:cmd}
\subsection{Spatial Selection of Sources}
\begin{figure}
\begin{center}
\includegraphics[width=0.48\textwidth]{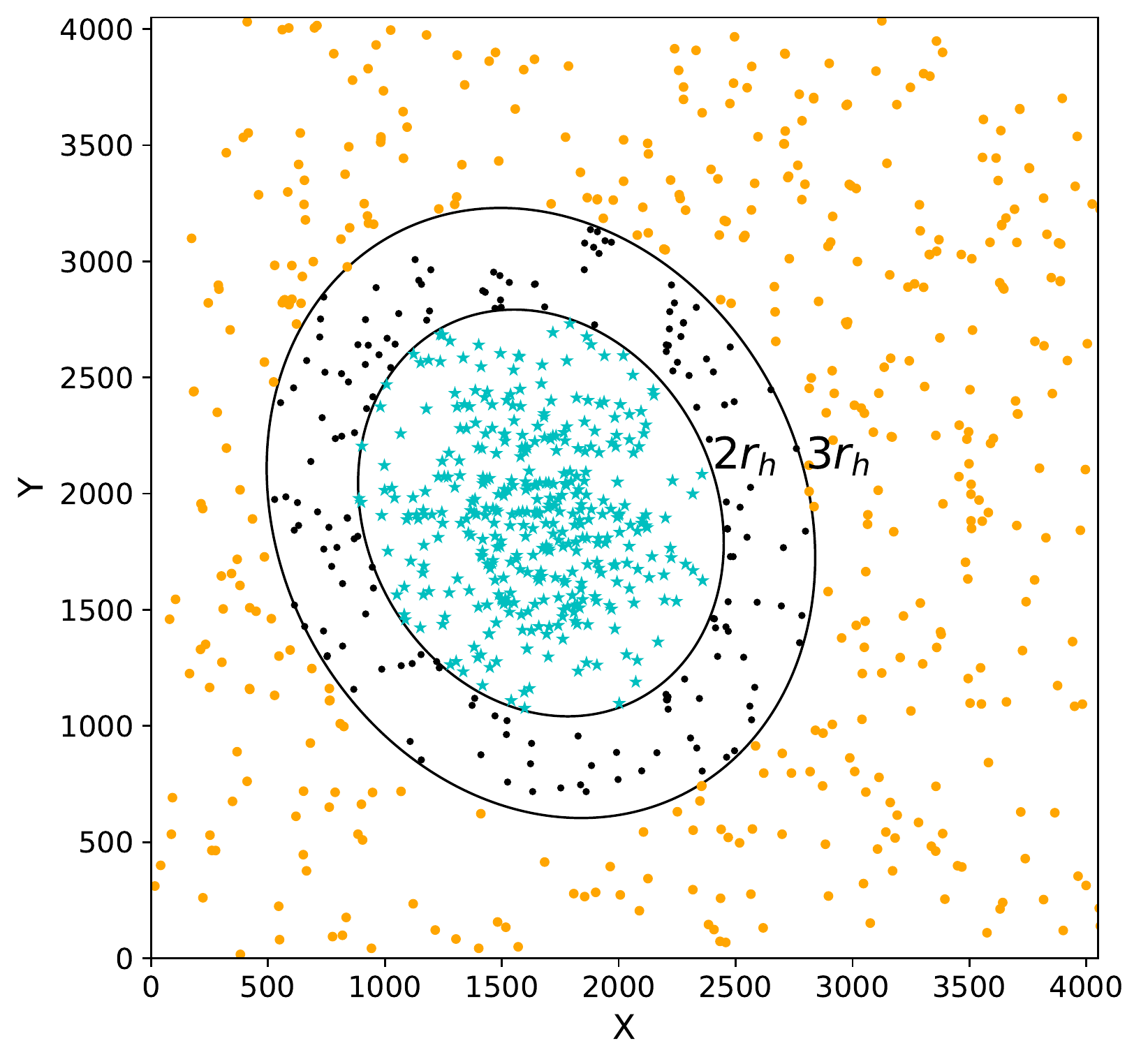}
\end{center}
\caption{Distribution of point sources passing the photometric quality cuts in the X--Y coordinates of the ACS field of view. \peg\ is clearly identifiable as an over density of point sources. The smaller ellipse is based on our measured structural parameters of \peg\, extends to $2~r_h$, and encircles the point sources (cyan stars) used in our analysis of the galaxy. The larger ellipse reaches $3~r_h$; sources outside this area (orange circles) quantify the potential contamination of the final \peg\ stellar catalog. The sources located between $2< r_h < 3$ (black points) are not used in our analysis.}
\label{fig:spatial}
\end{figure}

Figure~\ref{fig:spatial} presents the spatial distribution of all sources that pass our photometric quality cuts in X-Y coordinates. Two ellipses are overplotted based on our structural parameters with semi-major axes of $2~r_h$ and $3~r_h$, respectively. The overdensity of sources from \peg\ is readily apparent in the smaller ellipse. Outside of this region there are still a few hundred sources that pass our quality cuts. While some of the sources may be bona fide members of \peg, the majority are likely foreground stars or unresolved background galaxies. 

Thus, to create a high-fidelity photometric catalog of \peg\ with minimal contamination, we used the structural parameters and applied spatial cuts to the data. Our final stellar catalog for \peg\ includes all sources contained within the ellipse extending to $2~r_h$, shown as cyan stars in Figure~\ref{fig:spatial}, and includes 377 sources. To quantify the potential contamination to the \peg\ catalog, we use the sources outside the larger ellipse that extends to $3~r_h$ and create a field catalog. These sources are shown as orange circles and include a total of 465 sources from an area $\sim\,7\times$ larger than that used for \peg. This field catalog is used to estimate contamination in the CMD in Section~\ref{sec:cmd} and to model the contamination in the SFH fitting in Section~\ref{sec:sfh}. The sources in between $2-3r_h$ are not used in either catalog and are shown as black circles. 

Note that the WFC3 UVIS parallel field was originally intended to quantify the potential contamination of the final \peg\ stellar catalog. We examined the WFC3 data and found a somewhat smaller number of sources (241) passed the photometric quality cuts from the full UVIS field of view. We opt to use the field catalog from the ACS data as representative of the potential contamination as the field is closer spatially to \peg, avoids transforming from WFC3 UVIS to ACS magnitudes, and also avoids any systematics caused by slightly different spatial resolution and throughputs.

\subsection{CMD of Final Stellar Catalog}
\begin{figure*}
\begin{center}
\includegraphics[width=0.98\textwidth]{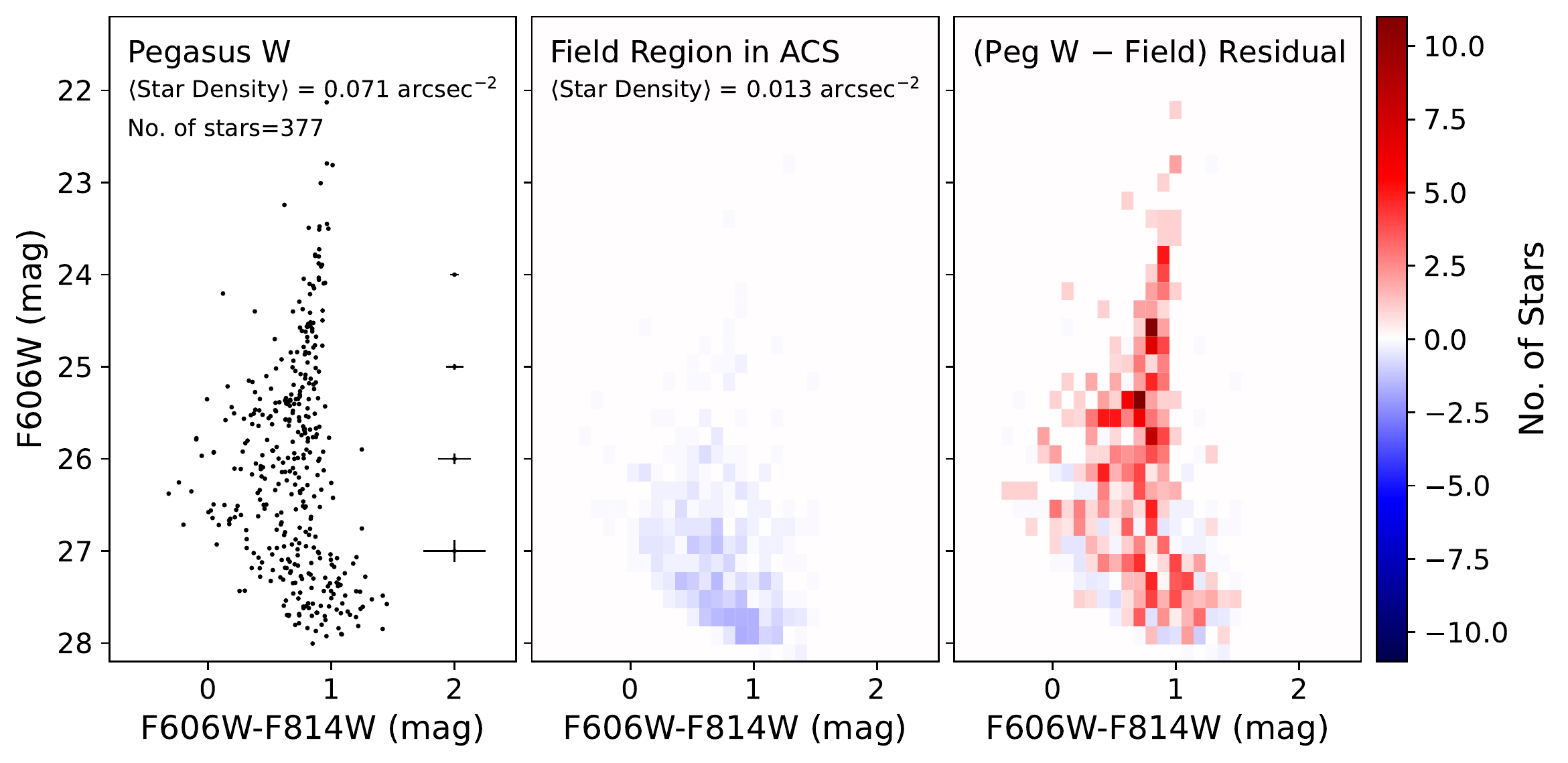}
\end{center}
\caption{Left: CMD of the 377 sources located inside an ellipse centered on \peg\ and extending to $2~r_h$. The CMD is plotted to the $\sim$\,50\% completeness limit; representative uncertainties per magnitude are shown. Middle: Hess diagram of sources in the ACS field of view outside $3~r_h$ that passed our quality cuts, scaled to the area encompassing the \peg\ region. Right: Residual Hess diagram made from subtracting the field scaled Hess diagram from the \peg\ data. Red indicates a higher number of sources in \peg\ while blue indicates a higher amount of contamination.}
\label{fig:cmds}
\end{figure*}

Figure~\ref{fig:cmds} shows the CMD from the final photometric catalog for \peg\ in the left panel, a 2D histogram representation of the CMD (i.e., a Hess diagram) of the field catalog scaled to the same area as \peg\ in the middle panel, and a residual Hess diagram of the field CMD subtracted from the \peg\ CMD. The stellar density of the \peg\ catalog is 0.072 sources per arcsec$^2$, compared with 0.013 in the field catalog. The data are plotted to the 50\% completeness limits and representative photometric uncertainties per magnitude bin are shown in the \peg\ CMD; no corrections for completeness have been applied to the stellar density values.

The CMD for \peg\ shows well-defined RGB, a blue and red horizontal branch (HB), and red clump (RC) features, which are largely absent in the field Hess diagram. The differences between the CMDs of \peg\ and the field region can be seen more quantitatively in the residual Hess diagram in the right panel of Figure~\ref{fig:cmds}. The red color represents higher star counts from \peg\ over the field region and includes the RGB and HB features in the CMD. The blue color indicates that contamination dominates the source counts and is seen predominantly at fainter magnitudes.

There are also two groups of sources in the CMD that warrant additional attention. There are four sources blue-ward of the RGB and above the HB feature (i.e., F606W$-$F814W $<0.8$; F606W $<$24.8) which have optical colors and magnitudes consistent with blue helium burning (BHeB) stars with ages of $<$\,500 Myr \citep{McQuinn2011}. There is one source in a similar region of CMD-space in the catalog from the off-galaxy field area that is $\sim\,7\times$ larger. 

There is also a grouping of  thirty blue sources fainter than the HB with colors $\ltsimeq 0.1$ mag that could be the bright end of what is referred to as the `blue plume'. The blue plume is often detected in the CMDs of globular clusters and UFDs that reach the old main sequence turn-off \citep[oMSTO; e.g.,][]{Momany2007, Mapelli2007, Mapelli2009, Sand2010, Monelli2012, Okamoto2012, Santana2013}. The nature of these sources is unclear. They may be young ($\ltsimeq3$ Gyr) main sequence stars (possibly with unresolved binaries), indicating that a galaxy has hosted late-time star formation. They could also be blue stragglers (BSs), which are older, hot, blue stellar sources with optical colors and magnitudes similar to younger main sequence stars \citep{Mateo1995}. BSs are thought to be a product of stellar binary systems, with a collisional or mass-transfer origin. 

In the case of \peg, the population of faint, blue stars lie at the bright end of what is typically considered the blue plume region identified in CMDs reaching deeper photometric depths. The residual Hess diagram in the final panel of Figure~\ref{fig:cmds} reveals this is a mixed population of high-confidence \peg\ member stars and contamination. Specifically, there are an average of 11 blue plume sources in a region of equal area to \peg\ compared with 30 found within $2~r_h$ of the galaxy. The BHeB star candidates and the blue plume sources are both consistent with ages of $\sim$500 Myr, based on BaSTI isochrones \citep{Hidalgo2018}. Examining the spatial distribution shows that 2 of the 4 BHeB candidates and 10 of the 30 blue plume sources are concentrated in the inner $r_h$, with 4 blue plume sources within $0.5~r_h$. BSs in dwarf galaxies are expected to have a flat radial distribution whereas younger stars are more spatially clumped \citep[e.g.,][]{Momany2007, Monelli2012}, lending support to the idea that some of the sources may be young stars. 

The specific fraction of blue stragglers to RGB stars has been shown to be approximately constant in UFD galaxies \citep{Santana2013}. Thus, to further investigate the nature of the blue plume sources, we follow the procedure in \citet{Santana2013} and calculate the fraction of blue plume to RGB stars after subtracting the number of sources in the field region scaled by the areal coverage of the galaxy vs.\ field region. Note, however, that the specific fraction we calculate is a lower limit on how many blue stragglers are expected. Our data are in different filters and do not reach the same photometric depths as the data from \citet{Santana2013}. As a result, the magnitude ranges used to select the population of stars are not identical and, given the relative faintness of blue plume to RGB stars, this has a greater impact on the number of blue plume stars counted. Our calculation yields a lower limit of the specific fraction of 0.21 $-$ below the average value of 0.29$\pm$0.1 from \citet{Santana2013} $-$ making our results inconclusive on the nature of the blue plume sources using this metric. Note, an Anderson-Darling statistical test performed on the cumulative radial distributions of the blue plume and galaxy population samples was also inconclusive. 

The galaxy was not detected in GALEX near or far UV imaging, although this is to be expected given the shorter star formation timescale that UV emission traces and the depth of the GALEX data. \peg\ was also not detected in the \hi\ ALFALFA survey \citep{Haynes2018}, but it is possible that the galaxy still harbors some gas below the detection limit of the survey.

While it is intriguing that \peg\ may have hosted late-time star formation, this is still speculative as it is not possible to discern with confidence the true nature of the BHeB candidates or blue plume sources without deeper imaging or spectroscopic data. We return to this when fitting for the SFH in Section~\ref{sec:sfh}.

\section{The Distance to Pegasus~W}\label{sec:distance}
The distance to the galaxy was measured to be 915$^{+60}_{-91}$ kpc based on the luminosity of the HB stars. We chose to use the HB stars for a standard candle distance determination, instead of the tip of the red giant branch stars, as the upper RGB is sparsely populated and the brightest identified stars in the RGB sequence may be a false tip \citep[cf.][]{McQuinn2013, McQuinn2015a}.

\begin{table*}
\begin{center}
\caption{Photometry Transformation Coeffcients}
\label{tab:transform}
\end{center}
\begin{center}
\vspace{-15pt}
\begin{tabular}{lrr | rr}
\hline 
\hline 
Coeff.		& \multicolumn{2}{c}V			& \multicolumn{2}{c}I \\
		& $V-I<0.4$		& $V-I\ge0.4$		& $V-I<0.1$			& $V-I<0.1$ \\
\hline
$c_0$		& 26.394$\pm0.05$	& 26.331$\pm0.008$	& 25.479$\pm0.13$		& 25.496$\pm0.010$	\\
$c_1$		& 0.153$\pm0.018$	& 0.340$\pm0.008$		& 0.041$\pm0.211$		& $-0.014\pm0.013$		\\
$c_2$		& 0.096$\pm0.085$	& $0.038\pm 0.002$		& $-0.093\pm0.803$		& 0.015$\pm0.003$		\\
\hline
\hline              
\end{tabular}
\end{center}
\tablecomments{Coefficients used in Eq.~\ref{eq:transform} for transforming the ACS photometry to Johnson $V$ and $I$ magnitudes as a function of $V-I$  color.}
\end{table*}

The HB distance methodology arises from the nearly constant $V$-band luminosity of helium burning stars in the post-RGB phase of evolution. The absolute magnitude of the HB has been calibrated in the Johnson BVRI filter system. Thus, we convert the photometry from the HST ACS filters to the Johnson filter system using the transformations from \citet[][see their Equation 12 with coefficients from Table 22]{Sirianni2005}, and iteratively solve for $V$ and $I$ band magnitudes using the F606W$-$F814W colors as a starting point:

\begin{equation}
\begin{split}
V = F606W + c_0 + c_1 * (V - I) + c_2 * (V - I)^2 \\
I = F814W + c_0 + c_1 * (V - I) + c_2 * (V - I)^2 
\end{split}
\label{eq:transform}
\end{equation}

\noindent The coefficients $c_0$, $c_1$, $c_2$  are given in Table~\ref{tab:transform}. Before transforming to the $V,I$ magntiudes, the photometry was corrected for Galactic foreground extinction using the dust maps from \citet{Schlegel1998} and the re-calibration by \citet{Schlafly2011}; values are listed in Table~\ref{tab:properties}. 

The HB luminosity was measured using a maximum likelihood approach to fitting a parametric luminosity function to the stars in the region of the HB with the following form:
\begin{equation}
P = e^{A(V-V_\text{HB})+B} + e^{-0.5[(V-V_\text{HB})/C]^2},
\label{eq:hb_parameteric}
\end{equation}
where $V$ is the magnitude in the F606W filter, and $A$, $B$, and $C$ are free parameters. The parameter $C$ has a prior of $\frac{1}{C}$. An advantage of using a maximum likelihood approach is that it takes into account the distribution of photometric uncertainties and completeness measured by the artificial star tests. 

\begin{figure}
\begin{center}
\includegraphics[width=0.35\textwidth]{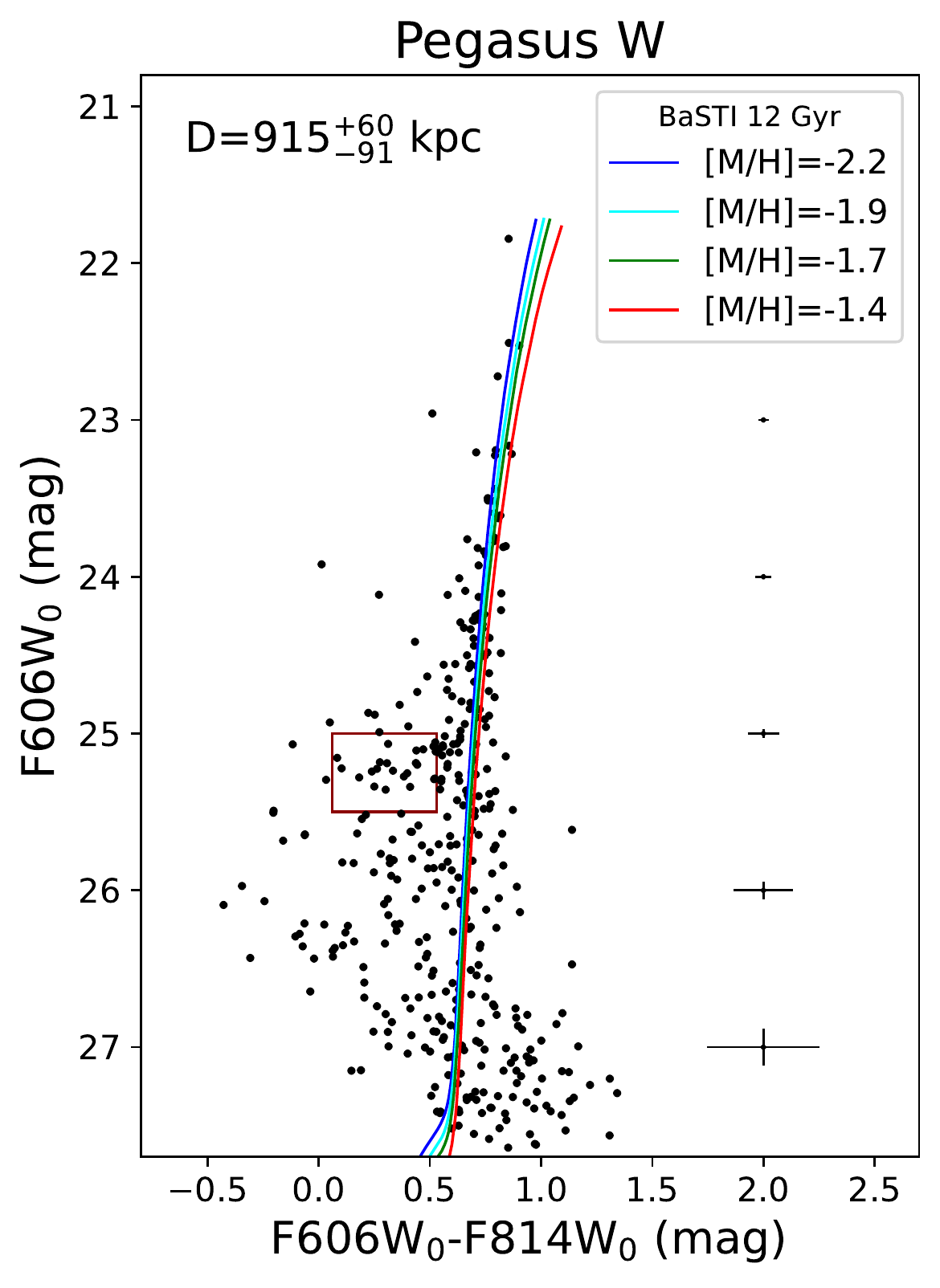}
\end{center}
\caption{Extinction corrected CMD of \peg\ with 12 Gyr BaSTI isochrones and varying [M/H] values overlaid. The red box outlines the area of the CMD used to fit for the HB luminosity.}
\label{fig:cmd_isochrone}
\end{figure}

Figure~\ref{fig:cmd_isochrone} presents the extinction corrected CMD for \peg. Stars used for the fit lie in the range of the HB identified by eye in the CMD, as shown by the red box in Figure~\ref{fig:cmd_isochrone}, which avoids including stars on the blue edge of the red clump. The best-fit HB luminosity is $V_0 = 25.30^{+0.12}_{-0.20}$ mag.

The HB luminosity was converted to a distance modulus using the calibration from \citet{Carretta2000}:
\begin{equation}
M_\text{V} = (0.13\pm 0.09) \times {\rm [Fe/H] + 1.5} + (0.54\pm0.07)
\label{eq:hb_calibrate}
\end{equation}
\noindent We estimate [Fe/H] by overlaying isochrones on the CMD. Shown in Figure~\ref{fig:cmd_isochrone} are a set of 12 Gyr BaSTI isochrones with [M/H] ranging from $-2.2$ to $-1.4$. Based on the CMD, we find that the [M/H] $=-$1.9 and $-$2.2 are the best overall matches to the RGB. The higher value of [M/H] $=-1.9$ is in agreement with the best-fitting average stellar metallicity found from the CMD-fitting technique using the BaSTI stellar library (see Section~\ref{sec:sfh}). Thus, we adopt [M/H] = $-1.9$ as representative of [Fe/H] in the distance calculation with an uncertainty of 0.1 dex. This metallicity estimate is somewhat higher than typically reported for UFD galaxies, but not unexpected given the extended SFH presented in the next section.

\begin{figure*}
\begin{center}
\includegraphics[width=0.98\textwidth]{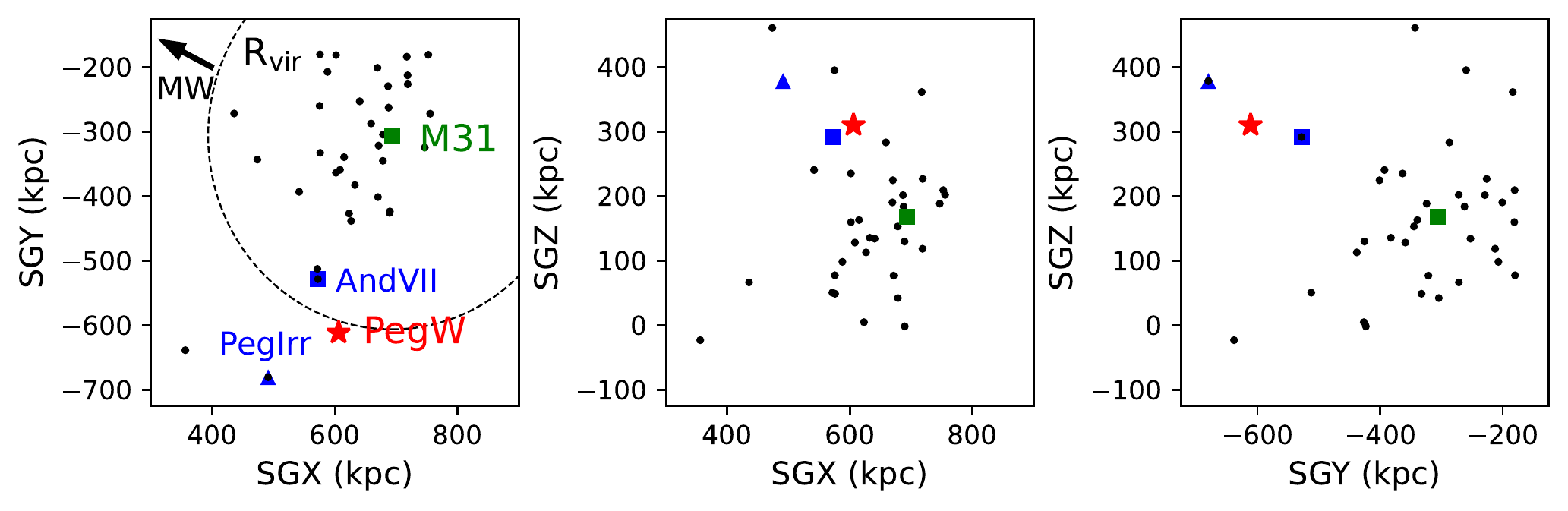}
\end{center}
\caption{Known systems in SG coordinates within 500 kpc of \peg\ based on the updated compilation of galaxies from \citet{McConnachie2012} and revised distances for a subset from \citet{Savino2022}. The location of \peg\ is shown as a red star; the uncertainties in the position of the galaxy based on distance uncertainties are smaller than the plot symbol. M31 is shown as a green square; the majority of black points are satellites of M31. The nearest known neighbors to \peg\ are And~VII (blue square) and the Pegasus dwarf galaxy (blue triangle). The MW is located at the SG origin on the far side of M31 from \peg; the black arrow in the first panel indicates the direction of the MW in the SGX$-$SGY plane. }
\label{fig:SG}
\end{figure*}

The final distance modulus to \peg\ is 24.81$^{+0.14}_{-0.22}$ mag, corresponding to a distance of 915$^{+60}_{-91}$ kpc. The uncertainties are based on adding in quadrature the uncertainties in the HB luminosity fit, the estimated uncertainty on the adopted [Fe/H] value, and the uncertainties from the calibration in Eq.~\ref{eq:hb_calibrate}.

\subsection{Location within the Local Group}
A secure distance enables the location of \peg\ within the Local Group to be firmly established. Figure~\ref{fig:SG} shows the distribution of sources within a physical 3-dimensional distance of 500 kpc of \peg\ in Supergalactic (SG) euclidian coordinates (SGX, SGY, SGZ), which provide a way to visualize the physical spatial distribution of galaxies and their linear separation. The SG coordinates were determined using the spherical coordinates and distance to each system following the formalism described in \citet{McQuinn2014}. For reference, the MW sits at the SG origin. Note that the axis ranges on each plot are fixed at 600 kpc, but the plot includes only galaxies found to be within 500 kpc of \peg\ in the updated compilation from \citet{McConnachie2012} with revised distances from \citet{Savino2022} were available. We also include M31 with an adopted distance of 776.2$^{+22}_{-21}$ kpc determined from RR Lyrae stars \citep{Savino2022}.

From Figure~\ref{fig:SG}, \peg\ (red star) lies on the far side of the MW$-$M31 system at a distance of 348 kpc from M31, which places it outside an assumed 300 kpc virial radius (green square). The galaxy is relatively isolated;  the two nearest neighbors are the M31 satellite, AndVII (separation$=$92 kpc; blue square) and the low-mass galaxy Peg Irr (separation$=$150 kpc; blue circle). The clustering of black points are comprised mainly of satellites of M31. 

\section{The Star Formation History of \peg}\label{sec:sfh}
\subsection{SFH Methodology}
The SFH of the galaxy was reconstructed from the stellar populations in the CMD using the CMD-fitting software {\tt MATCH} \citep{Dolphin2002}. The basic approach of a CMD-fitting technique is to find the best-fitting combination of synthetic simple stellar populations (SSPs) of different ages and metallicities from stellar evolution libraries to an observed CMD. The SSPs are constructed using an assumed initial mass function (IMF) and binary fraction and then linearly combined until the best-fit is found based on a Poisson likelihood statistic.

For the SFH fits for \peg, we assumed a Kroupa IMF \citep{Kroupa2001}, a binary fraction of 35\% with flat secondary mass distribution, and used as primary inputs the photometry and observational uncertainties and incompleteness measured from the artificial star recovery fractions. The SFHs were reconstructed using three stellar evolution libraries, namely BaSTI, PARSEC \citep{Bressan2012}, and MIST \citep{Choi2016}, which help to bracket the range of possible solutions and provide an estimate of the systematic uncertainties. Given the photometric depth of the data, the CMD does not fully constrain the metallicity evolution of \peg. Thus, we imposed a physically motivated constraint that the metallicity is a continuous, non-decreasing function over the lifetime of the galaxy. The SFH solutions were fit with an age grid of $\log(t/\text{yr}) = 6.6$--10.15 using a time resolution $\delta t=0.1$ dex for ages less than $\log(t/\text{yr})=9.0$ and $\delta t=0.05$ dex for older ages, and a metallicity grid [M/H]$=-2.0$ to $-1.0$ with a resolution of 0.15 dex. We initially tested searching a metallicity grid encompassing lower metallicity values, but as the best-fitting [M/H] solutions were consistently above $-2.0$, we tightened the search grid to reduce computational time. We fixed the distance in the fits to the HB distance of 0.915 Mpc and adopt a Galactic extinction value of $A_V=0.315$ mag based on the dust maps from \citet{Schlegel1998} and recalibration from \citet{Schlafly2011}. 

\begin{figure}
\begin{center}
\includegraphics[width=0.48\textwidth]{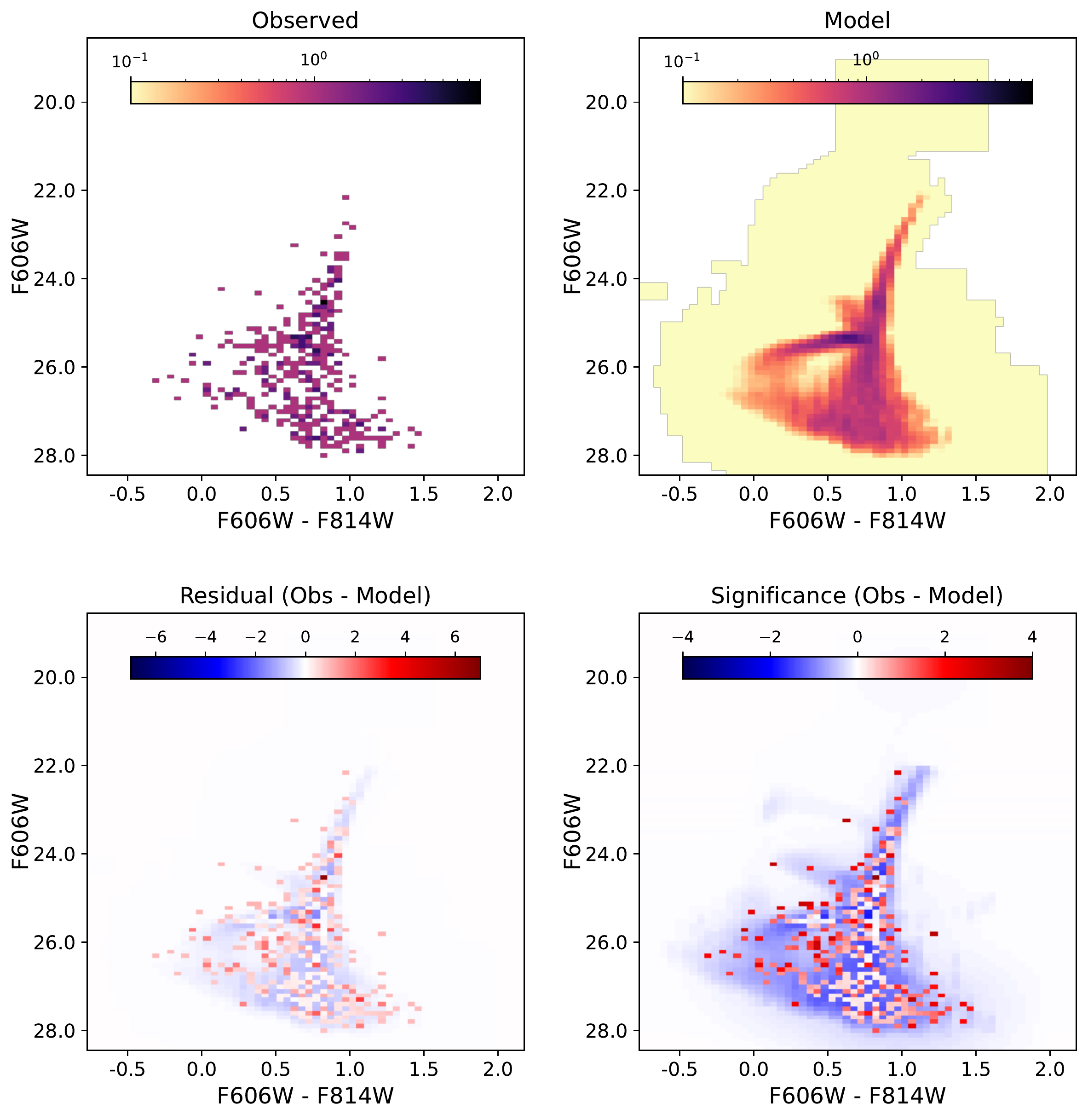}
\end{center}
\caption{Example of the quality of fit to the observed CMD using the BaSTI stellar library. Top left: observed CMD; top right: modelled Hess diagram used in the fit; bottom left: simple residual Hess diagram (data $-$ model); bottom right: residual significance Hess diagram where the pixels are weighted by the variance. The checkerboard pattern in the residual significance indicates there are no major residuals and the modelled Hess diagram is a good fit to the data.}
\label{fig:residuals}
\end{figure}

Figure~\ref{fig:residuals} shows an example of the quality of the fit to the data using the BaSTI stellar library. The most important diagnostic is the residual significance plot (bottom right panel;  units in standard deviations); a checkerboard pattern of blue and red indicates no major residuals. The overall fit is quite good and the different evolutionary sequences are well reproduced by the model.

Random uncertainties due to the finite number of stars in a CMD were estimated using a Markov Chain Monte Carlo approach \citep{Dolphin2013}. In addition, while the SFH derived using the different stellar libraries provides a measure of the range of possible solutions, systematic uncertainties were also estimated using 50 Monte Carlo simulations \citep{Dolphin2012}.

\subsection{Best-Fitting SFH}
The left panel of Figure~\ref{fig:sfh} presents the best-fitting SFH using the BaSTI models; the uncertainties include both random and systematic uncertainties. The right panel shows the best-fitting SFH solutions with BaSTI, MIST, and PARSEC models with random uncertainties only.  The SFH recovered by the three stellar libraries are in overall good agreement. The range in solutions from the different libraries lie within the systematic uncertainties estimated for the BaSTI library shown in the left panel. For the downstream analysis, we adopt the SFH solution based on the BaSTI library which produced a slightly better fit overall than the PARSEC and MIST libraries. 

The best-fitting SFH for \peg\ shows a rising star formation until $\sim\,7$ Gyr ago, with an indication of late-time star formation activity over the past few Gyr. As discussed above, there are a few BHeB star candidates and sources in blue plume region of the CMD that may be bona fide young stars. If these are true BHeB and young main sequence stars, they can provide valuable and unique constraints on late-time star formation in an UFD. On the other hand, the BHeB candidates could be contaminants in the CMD and the blue plume sources could be BSs masquerading as young stars. To try to account for the possibility that BHeB are contaminants, we use the field catalog as representative of potential background sources while fitting for the SFH. However, BSs are currently not included in the stellar evolution libraries used in the CMD-fitting and, thus, cannot be explicitly taken into account in the SFH recovery. In this case, including them in the fit will falsely result in star formation activity $\ltsimeq3$ Gyr ago. 

To explore the impact of the candidate BHeB stars and blue plume on our results, we fit for the SFH with and without these sources. Excluding these sources has little impact on the total stellar mass recovered from the data, reducing the stellar mass formed in the galaxy by $\sim$\,1--2\% depending on the stellar library. Removing these sources eliminates star-formation activity $<500$ Myr ago, but not the low level of star formation 1--3 Gyr ago. We include these sources in our final fit, but note there is a larger uncertainty in the very recent SFH ($t<500$ Myr) given the ambiguity of the nature of these sources. 

\subsection{Star-Formation Timescales}
Of particular interest is the timescale by which the star formation ceased, or the quenching timescale. We adopt the lookback time at which the galaxy has formed 90\% of its stellar mass as the quenching timescale (\tninety). Using \tninety\ reduces the impact blue plume stars have on quenching timescale as it is not as dependent on whether the galaxy has experienced a small level of recent star formation activity.

\begin{figure}
\begin{center}
\includegraphics[width=0.48\textwidth]{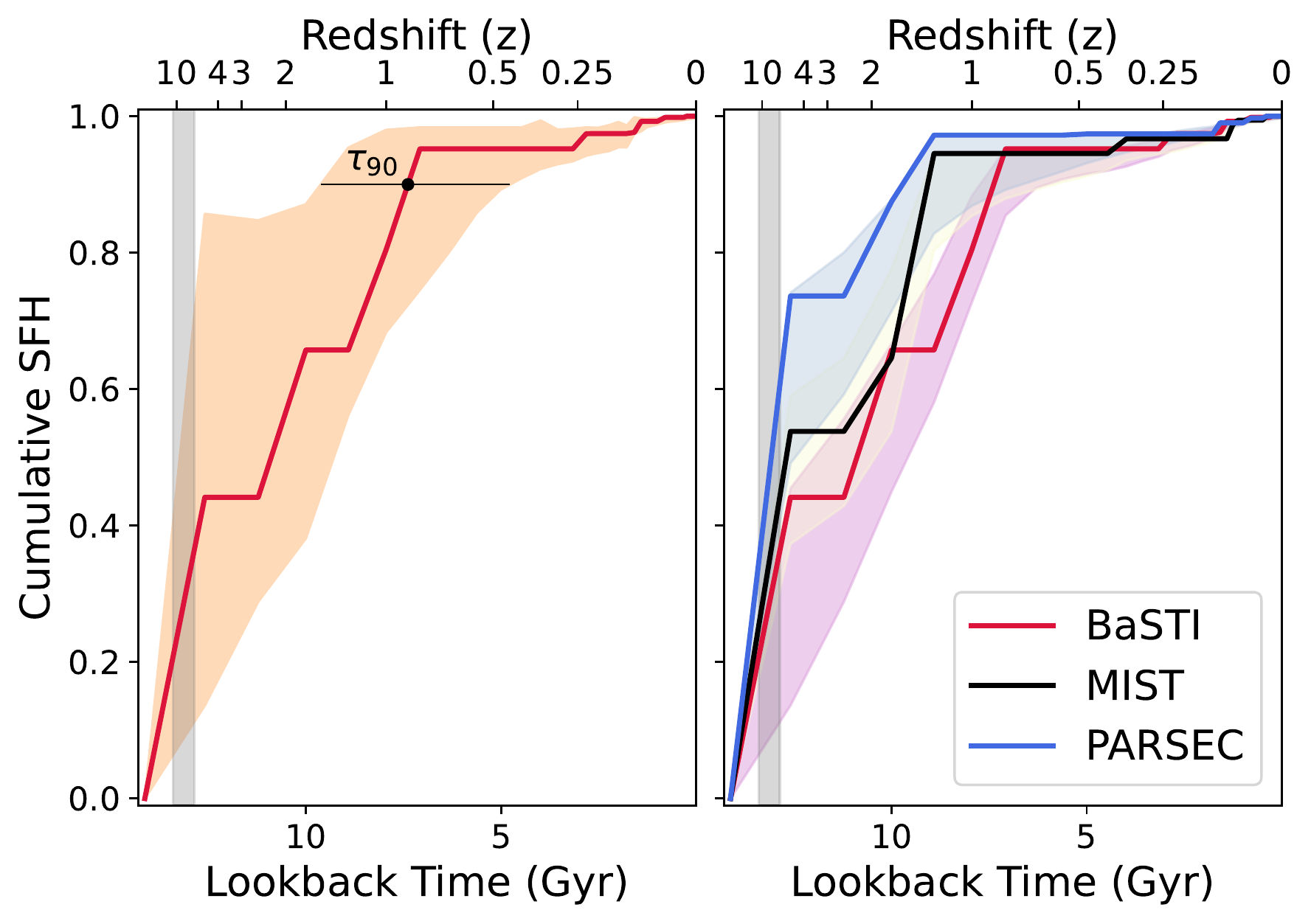}
\end{center}
\caption{Left: Best-fitting SFH solution with the BaSTI models. Uncertainties include both statistical and systematic uncertainties. The quenching timescale, \tninety\ is marked as noted. Right: Best-fitting SFH solutions using the BaSTI, MIST, and PARSEC models with statistical uncertainties only. The gray shaded vertical region shows the approximate timescale of reionization.}
\label{fig:sfh}
\end{figure}

We find \tninety\ $=7.4^{+2.2}_{-2.6}$Gyr based on interpolating the SFH derived using the BaSTI stellar library. The uncertainties include both statistical and systematic uncertainties shown in the left panel of Figure~\ref{fig:sfh} which were similarly determined by interpolating the uncertainty envelopes. The value of \tninety\ with uncertainties is overplotted on the SFH in the left panel of Figure~\ref{fig:sfh}. 

\subsection{Total Luminosity of \peg}
We calculate the present day $M_{V}$ of \peg\ following the approach outlined in \citet{Martin2008}. To account for uncertainties in both distance (resulting from our fit to the HB magnitude) and the structural parameters, we perform a Monte Carlo procedure.  Specifically, in each Monte Carlo iteration, we generate a synthetic stellar population using the best-fit star formation history, a distance drawn from the HB-based distance determination, and a value of $N_{\star}$ from the posterior distribution in the previous section. Since $N_{\star}$ represents the total number of stars in the system \textit{observed within our chosen CMD limits} of F606W $<$ 26.6, F814W $<$ 25.7 and $(\textrm{F606W} - \textrm{F814W})$ $<$ 1.5, we continue to draw stars from the \textit{entire} synthetic population until the number of stars that would be observed within these CMD limits is equal to $N_{\star}$.  In each iteration, the F606W magnitudes of all stars drawn from the synthetic CMD are corrected for extinction and distance and converted to $V$-band magnitudes using the bolometric corrections of \citet{Chen2019} assuming [M/H] $=-1.9$, and their sum then yields a total $M_{V}$ in each iteration.  The value and uncertainties we report correspond the median and 16th$-$84th percentiles obtained over all iterations, yielding $M_{V} = -7.20^{+0.17}_{-0.16}$ mag.  

We also attempted to directly estimate the total luminosity of \peg\ using publicly available images from the DESI Legacy Imaging Surveys \citep{Dey2019}. We obtained an estimate that is consistent with Monte Carlo method; however, the uncertainty on the estimate is much larger ($\gtsimeq\,$0.5\,mag) due to the image's limited depth.

\subsection{Present-Day Stellar Mass}
We estimate the present-day stellar mass, \mstar, in \peg\ using two approaches. First, we  obtain an estimate by using the results of the Monte Carlo simulations used to determine the total luminosity the galaxy, described above. We sum the total mass of all stars drawn in each Monte Carlo iteration, which yields \mstar\ $= 6.9^{+1.1}_{-1.0} \times 10^4$ \msun.     

Second, \mstar\ is estimated directly from the SFH. The {\em total} stellar mass formed over the lifetime of the galaxy is determined while fitting for the SFH solution. This is akin to integrating the SFH over cosmic times, but is a separate quantity fit by assuming one time bin (i.e., $\log(t/\text{yr}) = 6.6$--10.15). The stellar mass determined in this way is equal to the value determined from integrating SFR(t), but the measurement avoids any co-variance between time bins in the fit and results in lower total uncertainties.  

To calculate the {\em present-day} stellar mass of \peg\, we make two adjustments. First, as described in \citet{Telford2020}, {\sc match} determines the stellar mass by integrating over a stellar mass range of $[0, \infty]$. This lower IMF normalization has the effect of systematically increasing the estimated stellar mass compared with a Kroupa IMF with typical mass limits of 0.1--100 \msun. Thus, we subtract 0.12 from log(\mstar/\msun) to account for the change in normalization, which lowers \mstar\ by 24\%. Second, we apply a recycling fraction, $R$, that estimates the fraction of material returned to the interstellar medium (ISM) from stars. We adopt a value of $R=0.41$ from \citet{Vincenzo2016}, based on a Kroupa IMF and the metallicity of \peg, and reduce the stellar mass by a factor of $1-R$. The final present-day stellar mass of \peg\ is found to be $6.1^{+0.9}_{-1.5}\times 10^4$ \msun\ and is in good agreement with the present-day stellar mass recovered from the SFH fits using the PARSEC and MIST stellar libraries. It is also in very good agreement with our first estimate based on the Monte Carlo simulations following the procedure in \citet{Martin2008}. We adopt an average of the two values and the largest uncertainties from both methods as our final value, yielding \mstar\ $=6.5^{+1.1}_{-1.5}\times10^4$ \msun.

\section{Discussion and Summary}\label{sec:discuss}
\begin{figure}
\begin{center}
\includegraphics[width=0.47\textwidth]{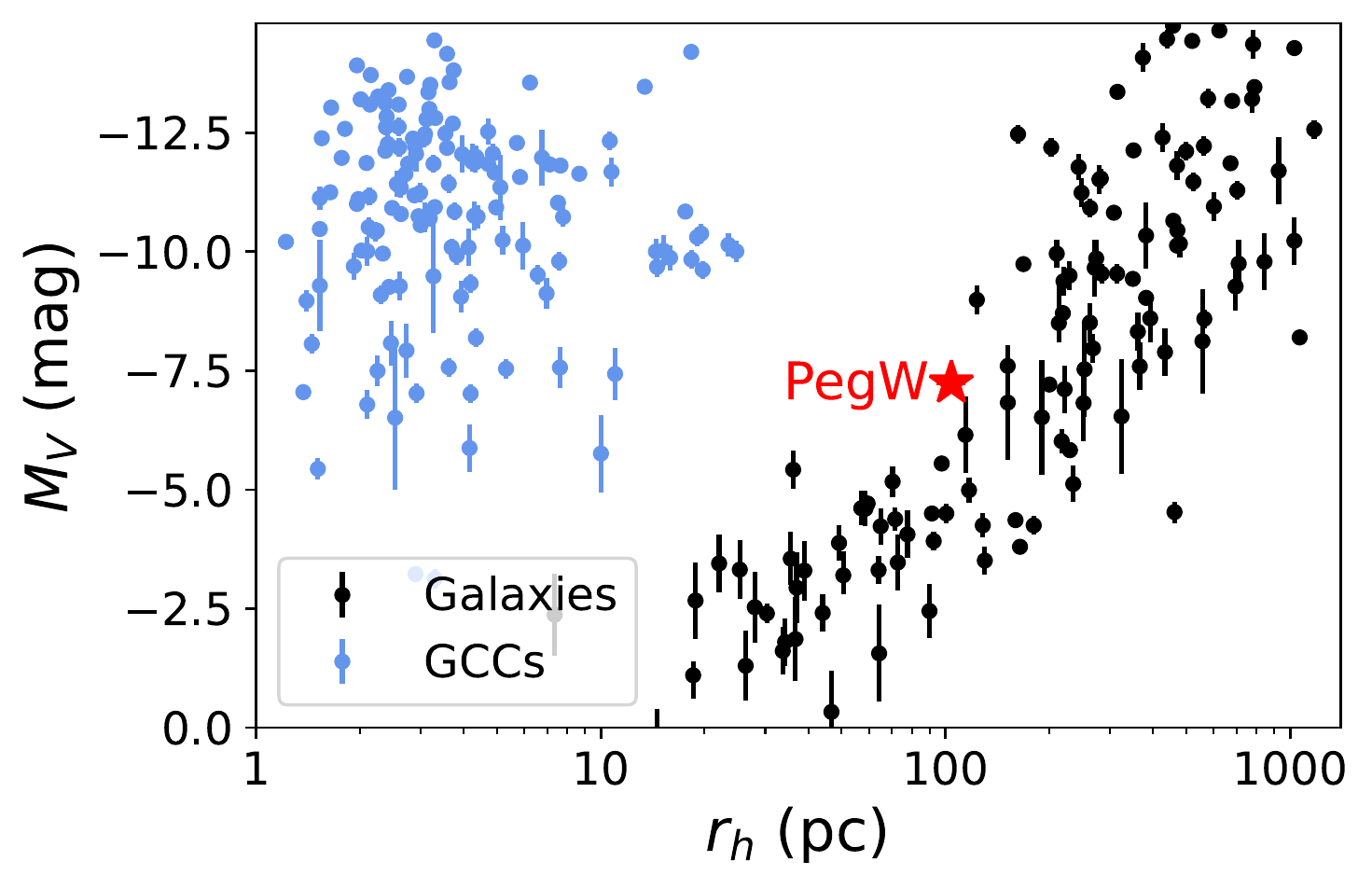}
\end{center}
\caption{Distribution of galaxies and Galactic globular clusters (GGCs) in the $M_V - r_h$ plane based on the updated compilation of galaxies from \citet{McConnachie2012} and \citet{Baumgardt2020}. Overlaid as a red star is the location of \peg\ using our measured parameters. The galaxy is consistent with the properties of known local dwarfs with comparable luminosity, although slightly more compact, and is considered an UFD galaxy based on the definition from \citet{Simon2019}. }
\label{fig:Mv_rh}
\end{figure}

\peg\ is a very low-mass (\mstar$=6.5^{+1.1}_{-1.4}\times10^4$ \msun), very faint ($M_V = -7.20^{+0.17}_{-0.16}$ mag) dwarf galaxy located on the far side of M31 at a distance of 915$^{+60}_{-91}$ kpc measured from HB stars. The galaxy lies outside the virial radius of M31 and is relatively isolated. Shown in Figure~\ref{fig:Mv_rh}, the size and luminosity of \peg\ are consistent with the properties of other local low-mass galaxies, including UFDs, although \peg\ is more compact than galaxies with similar luminosities. Based on the properties  of \peg\ reported here (distance, luminosity, half-light radius), the galaxy is an UFD in the Local Group. 

The best-fitting SFH derived from the CMD shows \peg\ has formed 10\% of its stellar mass within the last several Gyr, and suggests star-formation activity as recently as the last Gyr. We also identify BHeB candidate stars in the CMD that are consistent with age estimates $<500$ Myr. This late-time star formation is unusual in an UFD galaxy and indicates that the galaxy recently harbored gas, either through gas retention or recent accretion.

In addition to the evidence of late-time star formation, the overall SFH of \peg\ is extended: the galaxy formed 50\% of its stellar mass after $z \sim 6$ and quenched as recently as $z \sim 0.9$ (i.e., 7.4 Gyr ago). Cosmological simulations predict the mass threshold where reionization efficiently quenches galaxies to be \mstar\ $\sim 10^5$ \msun, possibly with an assist by stellar feedback \citep[e.g.,][]{Bovill2011, BenitezLlambay2015, RodriguezWimberly2019, Garrison-Kimmel2019, Rey2020, Wilson2022}. Above this threshold mass, a galaxy's potential should be deep enough to continue to accrete gas and the gas within a galaxy can self-shield sufficiently from the UV radiation to continue to cool and form stars. However, even though \peg\ has a stellar mass below this threshold (\mstar\ $6.5^{+1.1}_{-1.4}\times10^4$ \msun), the extended SFH is inconsistent with this expectation. Note that, while the stellar mass is directly observable, using \mstar\ as a measure of an UFDs total potential has large uncertainties given the spread in the stellar mass-halo mass relation of UFDs \citep[e.g.,][]{Munshi2021, Applebaum2021}. It's possible that despite having \mstar\ $<10^5$ \msun, the halo mass of \peg\ is large enough to be above the reionization threshold, although not significantly.

Interestingly, there is evidence that some of the UFD satellites of M31 in the mass range $10^4$ \msun\ $<$ \mstar\ $<10^5$ \msun\ also have somewhat extended SFHs and more recent quenching timescales. Initial results derived from data with a range of photometric depths (including And~XVI with imaging reaching below the oMSTO) found quenching timescales similar to what we find for \peg\ \citep{Skillman2017, Weisz2019}. Deeper observations for a subset of these systems place tighter constraints on the SFHs and shift the quenching timescales to slightly older lookback times  (Alessandro Savino \etal\ in preparation), but the timescales are still generally more recent than the rapid quenching by reionization scenario described above.

The more recent quenching timescale for \peg\ is in contrast to what has been derived for 6 UFDs of the MW in a slightly lower but overlapping mass range ($6\times10^3$ \msun\ $<$ \mstar\ $<5\times10^4$ \msun). These galaxies formed 80\% of their stars by $z\sim6$ (12.8 Gyr ago) and 100\% of their stars by $z\sim3$  \citep[11.6 Gyr ago;][]{Brown2014}. While the rapid quenching of the MW UFDs is consistent with and often attributed to reionization, it is possible that environment played a role in quenching these systems. Based on proper motions of MW UFDs from Gaia and modeling the time-varying gravitational potential of the MW due to its dark matter halo growth, there are indications that the SFHs of the UFDs were impacted by early environment effects from the MW halo \citep{Miyoshi2020}. If this is the case, then the differences in quenching timescales for the MW UFDs and \peg\ could be explained by differences in environment and orbital histories.

Indeed, rather than being quenched by reionization, it is more likely that \peg\ was quenched by environmental processing despite being outside the virial radius of M31. Based on the dwarf galaxy population around the MW and M31, there is a transition in the structural and dynamical properties of dwarfs at a distance of $\gtsimeq400$ kpc from their host (corresponding to $\sim1.33 \times R_\text{vir}$), which is thought to be due to tidal interactions \citep{Higgs2021}. Simulations have shown that tidal effects can extend well beyond the virial radius ($R_\text{vir}$) of the main halo \citep{Behroozi2014}, and that more massive dwarf galaxies (\mstar\ $\sim 10^{5.5}-10^8$ \msun) within 1--$2.5 \times R_\text{vir}$ of a more massive galaxy can be quenched by environment \citep{Fillingham2018}. It is also possible, even likely given \peg's current location at $1.2\times R_\text{vir}$, that the galaxy has previously passed through the halo of M31, making it a `backsplash' galaxy. Recent, extremely high-resolution cosmological simulations suggest that most galaxies within $1.5\times R_\text{vir}$ fall in the backsplash category \citep{Applebaum2021}. In this scenario, the galaxy would have experienced more significant environmental processing including ram pressure and tidal stripping \citep[e.g.,][]{Teyssier2012, Buck2019, Blana2020}. Such a fly-by interaction with M31 could have quenched the majority of star formation, but not completely stripped the \peg\ of gas. The remaining bound gas could have cooled over Gyr timescales, reigniting star formation and providing an explanation for the possible late-time star formation suggested in the CMD.

Note that the SFH of \peg\ is based on a CMD that is relatively deep, reaching $\sim\,2$ mag below the HB. However, that depth still falls meaningfully short of the oMSTO depth needed to break the age-metallicity degeneracy in the CMD at old lookback times. Our overall results would be improved should such data be acquired. In addition, spectroscopic observations of the stars in \peg\ could provide important constraints on the internal kinematics, dark matter content of the galaxy, and its current radial trajectory within the Local Group. 

\begin{acknowledgments}
Support for this work was provided by NASA through grant No.\ HST-GO-16916 and support for Y.-Y.M.\ was partly provided by NASA through the NASA Hubble Fellowship grant no.\ HST-HF2-51441.001, both were awarded by the Space Telescope Science Institute, which is operated by the Association of Universities for Research in Astronomy, Incorporated, under NASA contract NAS5-26555. K.B.W.M.\ is also supported by NSF grant AST-1940800. D.S.\ and M.R.B.\ are supported by DOE grant DOE-SC0010008. This research has made use of NASA Astrophysics Data System Bibliographic Services, adstex\footnote{https://github.com/yymao/adstex}, and the arXiv preprint server. 
\end{acknowledgments}

\facilities{Hubble Space Telescope}
\software{This research made use of {\tt DOLPHOT} \citep{Dolphin2000, Dolphin2016}, {\tt MATCH} \citep{Dolphin2002, Dolphin2012, Dolphin2013}, HST drizzlepac \citep[v3.0;][]{Hack2013, Avila2015}, {\tt emcee} \citep{Foreman-Mackey2013}, and Astropy,\footnote{http://www.astropy.org} a community-developed core Python package for Astronomy \citep{astropy:2018}.}

\renewcommand\bibname{{References}}
\bibliographystyle{apj}
\bibliography{ms.bbl}

\begin{thebibliography}{}
\expandafter\ifx\csname natexlab\endcsname\relax\def\natexlab#1{#1}\fi

\bibitem[{{Akins} {et~al.}(2021){Akins}, {Christensen}, {Brooks}, {Munshi},
  {Applebaum}, {Engelhardt}, \& {Chamberland}}]{Akins2021}
{Akins}, H.~B., {Christensen}, C.~R., {Brooks}, A.~M., {et~al.} 2021, \apj,
  909, 139

\bibitem[{{Applebaum} {et~al.}(2021){Applebaum}, {Brooks}, {Christensen},
  {Munshi}, {Quinn}, {Shen}, \& {Tremmel}}]{Applebaum2021}
{Applebaum}, E., {Brooks}, A.~M., {Christensen}, C.~R., {et~al.} 2021, \apj,
  906, 96

\bibitem[{{Avila} {et~al.}(2015){Avila}, {Hack}, {Cara}, {Borncamp}, {Mack},
  {Smith}, \& {Ubeda}}]{Avila2015}
{Avila}, R.~J., {Hack}, W., {Cara}, M., {et~al.} 2015, in Astronomical Society
  of the Pacific Conference Series, Vol. 495, Astronomical Data Analysis
  Software an Systems XXIV (ADASS XXIV), ed. A.~R. {Taylor} \& E.~{Rosolowsky},
  281

\bibitem[{{Baumgardt} {et~al.}(2020){Baumgardt}, {Sollima}, \&
  {Hilker}}]{Baumgardt2020}
{Baumgardt}, H., {Sollima}, A., \& {Hilker}, M. 2020, \pasa, 37, e046

\bibitem[{{Bechtol} {et~al.}(2015){Bechtol}, {Drlica-Wagner}, {Balbinot},
  {Pieres}, {Simon}, {Yanny}, {Santiago}, {Wechsler}, {Frieman}, {Walker},
  {Williams}, {Rozo}, {Rykoff}, {Queiroz}, {Luque}, {Benoit-L{\'e}vy},
  {Tucker}, {Sevilla}, {Gruendl}, {da Costa}, {Fausti Neto}, {Maia}, {Abbott},
  {Allam}, {Armstrong}, {Bauer}, {Bernstein}, {Bernstein}, {Bertin}, {Brooks},
  {Buckley-Geer}, {Burke}, {Carnero Rosell}, {Castander}, {Covarrubias},
  {D'Andrea}, {DePoy}, {Desai}, {Diehl}, {Eifler}, {Estrada}, {Evrard},
  {Fernandez}, {Finley}, {Flaugher}, {Gaztanaga}, {Gerdes}, {Girardi},
  {Gladders}, {Gruen}, {Gutierrez}, {Hao}, {Honscheid}, {Jain}, {James},
  {Kent}, {Kron}, {Kuehn}, {Kuropatkin}, {Lahav}, {Li}, {Lin}, {Makler},
  {March}, {Marshall}, {Martini}, {Merritt}, {Miller}, {Miquel}, {Mohr},
  {Neilsen}, {Nichol}, {Nord}, {Ogando}, {Peoples}, {Petravick}, {Plazas},
  {Romer}, {Roodman}, {Sako}, {Sanchez}, {Scarpine}, {Schubnell}, {Smith},
  {Soares-Santos}, {Sobreira}, {Suchyta}, {Swanson}, {Tarle}, {Thaler},
  {Thomas}, {Wester}, {Zuntz}, \& {DES Collaboration}}]{Bechtol2015}
{Bechtol}, K., {Drlica-Wagner}, A., {Balbinot}, E., {et~al.} 2015, \apj, 807,
  50

\bibitem[{{Behroozi} {et~al.}(2014){Behroozi}, {Wechsler}, {Lu}, {Hahn},
  {Busha}, {Klypin}, \& {Primack}}]{Behroozi2014}
{Behroozi}, P.~S., {Wechsler}, R.~H., {Lu}, Y., {et~al.} 2014, \apj, 787, 156

\bibitem[{{Ben{\'\i}tez-Llambay} {et~al.}(2015){Ben{\'\i}tez-Llambay},
  {Navarro}, {Abadi}, {Gottl{\"o}ber}, {Yepes}, {Hoffman}, \&
  {Steinmetz}}]{BenitezLlambay2015}
{Ben{\'\i}tez-Llambay}, A., {Navarro}, J.~F., {Abadi}, M.~G., {et~al.} 2015,
  \mnras, 450, 4207

\bibitem[{{Benson} {et~al.}(2002){Benson}, {Frenk}, {Lacey}, {Baugh}, \&
  {Cole}}]{Benson2002}
{Benson}, A.~J., {Frenk}, C.~S., {Lacey}, C.~G., {Baugh}, C.~M., \& {Cole}, S.
  2002, \mnras, 333, 177

\bibitem[{{Bla{\~n}a} {et~al.}(2020){Bla{\~n}a}, {Burkert}, {Fellhauer},
  {Schartmann}, \& {Alig}}]{Blana2020}
{Bla{\~n}a}, M., {Burkert}, A., {Fellhauer}, M., {Schartmann}, M., \& {Alig},
  C. 2020, \mnras, 497, 3601

\bibitem[{{Bovill} \& {Ricotti}(2011)}]{Bovill2011}
{Bovill}, M.~S., \& {Ricotti}, M. 2011, \apj, 741, 18

\bibitem[{{Bressan} {et~al.}(2012){Bressan}, {Marigo}, {Girardi}, {Salasnich},
  {Dal Cero}, {Rubele}, \& {Nanni}}]{Bressan2012}
{Bressan}, A., {Marigo}, P., {Girardi}, L., {et~al.} 2012, \mnras, 427, 127

\bibitem[{{Brown} {et~al.}(2014){Brown}, {Tumlinson}, {Geha}, {Simon},
  {Vargas}, {VandenBerg}, {Kirby}, {Kalirai}, {Avila}, {Gennaro}, {Ferguson},
  {Mu{\~n}oz}, {Guhathakurta}, \& {Renzini}}]{Brown2014}
{Brown}, T.~M., {Tumlinson}, J., {Geha}, M., {et~al.} 2014, \apj, 796, 91

\bibitem[{{Buck} {et~al.}(2019){Buck}, {Macci{\`o}}, {Dutton}, {Obreja}, \&
  {Frings}}]{Buck2019}
{Buck}, T., {Macci{\`o}}, A.~V., {Dutton}, A.~A., {Obreja}, A., \& {Frings}, J.
  2019, \mnras, 483, 1314

\bibitem[{{Bullock} \& {Boylan-Kolchin}(2017)}]{Bullock2017}
{Bullock}, J.~S., \& {Boylan-Kolchin}, M. 2017, \araa, 55, 343

\bibitem[{{Bullock} {et~al.}(2000){Bullock}, {Kravtsov}, \&
  {Weinberg}}]{Bullock2000}
{Bullock}, J.~S., {Kravtsov}, A.~V., \& {Weinberg}, D.~H. 2000, \apj, 539, 517

\bibitem[{{Carretta} {et~al.}(2000){Carretta}, {Gratton}, {Clementini}, \&
  {Fusi Pecci}}]{Carretta2000}
{Carretta}, E., {Gratton}, R.~G., {Clementini}, G., \& {Fusi Pecci}, F. 2000,
  \apj, 533, 215

\bibitem[{{Cerny} {et~al.}(2022){Cerny}, {Mart{\'\i}nez-V{\'a}zquez},
  {Drlica-Wagner}, {Pace}, {Mutlu-Pakdil}, {Li}, {Riley}, {Crnojevi{\'c}},
  {Bom}, {Carballo-Bello}, {Carlin}, {Chiti}, {Choi}, {Collins},
  {Darragh-Ford}, {Ferguson}, {Geha}, {Mart{\'\i}nez-Delgado}, {Massana},
  {Mau}, {Medina}, {Mu{\~n}oz}, {Nadler}, {Olsen}, {Pieres}, {Sakowska},
  {Simon}, {Stringfellow}, {Vivas}, {Walker}, \& {Wechsler}}]{Cerny2022b}
{Cerny}, W., {Mart{\'\i}nez-V{\'a}zquez}, C.~E., {Drlica-Wagner}, A., {et~al.}
  2022, arXiv e-prints, arXiv:2209.12422

\bibitem[{{Cerny} {et~al.}(2023){Cerny}, {Simon}, {Li}, {Drlica-Wagner},
  {Pace}, {Mart{\'\i}nez-V{\'a}zquez}, {Riley}, {Mutlu-Pakdil}, {Mau},
  {Ferguson}, {Erkal}, {Munoz}, {Bom}, {Carlin}, {Carollo}, {Choi}, {Ji},
  {Manwadkar}, {Mart{\'\i}nez-Delgado}, {Miller}, {No{\"e}l}, {Sakowska},
  {Sand}, {Stringfellow}, {Tollerud}, {Vivas}, {Carballo-Bello},
  {Hernandez-Lang}, {James}, {Nidever}, {Castellon}, {Olsen}, {Zenteno}, \&
  {Delve Collaboration}}]{Cerny2022a}
{Cerny}, W., {Simon}, J.~D., {Li}, T.~S., {et~al.} 2023, \apj, 942, 111

\bibitem[{{Chambers} {et~al.}(2016){Chambers}, {Magnier}, {Metcalfe},
  {Flewelling}, {Huber}, {Waters}, {Denneau}, {Draper}, {Farrow}, {Finkbeiner},
  {Holmberg}, {Koppenhoefer}, {Price}, {Rest}, {Saglia}, {Schlafly}, {Smartt},
  {Sweeney}, {Wainscoat}, {Burgett}, {Chastel}, {Grav}, {Heasley}, {Hodapp},
  {Jedicke}, {Kaiser}, {Kudritzki}, {Luppino}, {Lupton}, {Monet}, {Morgan},
  {Onaka}, {Shiao}, {Stubbs}, {Tonry}, {White}, {Ba{\~n}ados}, {Bell},
  {Bender}, {Bernard}, {Boegner}, {Boffi}, {Botticella}, {Calamida},
  {Casertano}, {Chen}, {Chen}, {Cole}, {Deacon}, {Frenk}, {Fitzsimmons},
  {Gezari}, {Gibbs}, {Goessl}, {Goggia}, {Gourgue}, {Goldman}, {Grant},
  {Grebel}, {Hambly}, {Hasinger}, {Heavens}, {Heckman}, {Henderson}, {Henning},
  {Holman}, {Hopp}, {Ip}, {Isani}, {Jackson}, {Keyes}, {Koekemoer}, {Kotak},
  {Le}, {Liska}, {Long}, {Lucey}, {Liu}, {Martin}, {Masci}, {McLean}, {Mindel},
  {Misra}, {Morganson}, {Murphy}, {Obaika}, {Narayan}, {Nieto-Santisteban},
  {Norberg}, {Peacock}, {Pier}, {Postman}, {Primak}, {Rae}, {Rai}, {Riess},
  {Riffeser}, {Rix}, {R{\"o}ser}, {Russel}, {Rutz}, {Schilbach}, {Schultz},
  {Scolnic}, {Strolger}, {Szalay}, {Seitz}, {Small}, {Smith}, {Soderblom},
  {Taylor}, {Thomson}, {Taylor}, {Thakar}, {Thiel}, {Thilker}, {Unger},
  {Urata}, {Valenti}, {Wagner}, {Walder}, {Walter}, {Watters}, {Werner},
  {Wood-Vasey}, \& {Wyse}}]{Chambers2016}
{Chambers}, K.~C., {Magnier}, E.~A., {Metcalfe}, N., {et~al.} 2016, arXiv
  e-prints, arXiv:1612.05560

\bibitem[{{Chen} {et~al.}(2019){Chen}, {Girardi}, {Fu}, {Bressan}, {Aringer},
  {Dal Tio}, {Pastorelli}, {Marigo}, {Costa}, \& {Zhang}}]{Chen2019}
{Chen}, Y., {Girardi}, L., {Fu}, X., {et~al.} 2019, \aap, 632, A105

\bibitem[{{Choi} {et~al.}(2016){Choi}, {Dotter}, {Conroy}, {Cantiello},
  {Paxton}, \& {Johnson}}]{Choi2016}
{Choi}, J., {Dotter}, A., {Conroy}, C., {et~al.} 2016, \apj, 823, 102

\bibitem[{{Collins} {et~al.}(2022){Collins}, {Charles},
  {Mart{\'\i}nez-Delgado}, {Monelli}, {Karim}, {Donatiello}, {Tollerud}, \&
  {Boschin}}]{Collins2022}
{Collins}, M. L.~M., {Charles}, E. J.~E., {Mart{\'\i}nez-Delgado}, D., {et~al.}
  2022, \mnras, 515, L72

\bibitem[{{Dey} {et~al.}(2019){Dey}, {Schlegel}, {Lang}, {Blum}, {Burleigh},
  {Fan}, {Findlay}, {Finkbeiner}, {Herrera}, {Juneau}, {Landriau}, {Levi},
  {McGreer}, {Meisner}, {Myers}, {Moustakas}, {Nugent}, {Patej}, {Schlafly},
  {Walker}, {Valdes}, {Weaver}, {Y{\`e}che}, {Zou}, {Zhou}, {Abareshi},
  {Abbott}, {Abolfathi}, {Aguilera}, {Alam}, {Allen}, {Alvarez}, {Annis},
  {Ansarinejad}, {Aubert}, {Beechert}, {Bell}, {BenZvi}, {Beutler}, {Bielby},
  {Bolton}, {Brice{\~n}o}, {Buckley-Geer}, {Butler}, {Calamida}, {Carlberg},
  {Carter}, {Casas}, {Castander}, {Choi}, {Comparat}, {Cukanovaite}, {Delubac},
  {DeVries}, {Dey}, {Dhungana}, {Dickinson}, {Ding}, {Donaldson}, {Duan},
  {Duckworth}, {Eftekharzadeh}, {Eisenstein}, {Etourneau}, {Fagrelius},
  {Farihi}, {Fitzpatrick}, {Font-Ribera}, {Fulmer}, {G{\"a}nsicke},
  {Gaztanaga}, {George}, {Gerdes}, {Gontcho}, {Gorgoni}, {Green}, {Guy},
  {Harmer}, {Hernandez}, {Honscheid}, {Huang}, {James}, {Jannuzi}, {Jiang},
  {Joyce}, {Karcher}, {Karkar}, {Kehoe}, {Kneib}, {Kueter-Young}, {Lan},
  {Lauer}, {Le Guillou}, {Le Van Suu}, {Lee}, {Lesser}, {Perreault Levasseur},
  {Li}, {Mann}, {Marshall}, {Mart{\'\i}nez-V{\'a}zquez}, {Martini}, {du Mas des
  Bourboux}, {McManus}, {Meier}, {M{\'e}nard}, {Metcalfe},
  {Mu{\~n}oz-Guti{\'e}rrez}, {Najita}, {Napier}, {Narayan}, {Newman}, {Nie},
  {Nord}, {Norman}, {Olsen}, {Paat}, {Palanque-Delabrouille}, {Peng},
  {Poppett}, {Poremba}, {Prakash}, {Rabinowitz}, {Raichoor}, {Rezaie},
  {Robertson}, {Roe}, {Ross}, {Ross}, {Rudnick}, {Safonova}, {Saha},
  {S{\'a}nchez}, {Savary}, {Schweiker}, {Scott}, {Seo}, {Shan}, {Silva},
  {Slepian}, {Soto}, {Sprayberry}, {Staten}, {Stillman}, {Stupak}, {Summers},
  {Sien Tie}, {Tirado}, {Vargas-Maga{\~n}a}, {Vivas}, {Wechsler}, {Williams},
  {Yang}, {Yang}, {Yapici}, {Zaritsky}, {Zenteno}, {Zhang}, {Zhang}, {Zhou}, \&
  {Zhou}}]{Dey2019}
{Dey}, A., {Schlegel}, D.~J., {Lang}, D., {et~al.} 2019, \aj, 157, 168

\bibitem[{{Dolphin}(2016)}]{Dolphin2016}
{Dolphin}, A. 2016, {DOLPHOT: Stellar photometry}, Astrophysics Source Code
  Library, record ascl:1608.013, ascl:1608.013

\bibitem[{{Dolphin}(2000)}]{Dolphin2000}
{Dolphin}, A.~E. 2000, \pasp, 112, 1383

\bibitem[{{Dolphin}(2002)}]{Dolphin2002}
---. 2002, \mnras, 332, 91

\bibitem[{{Dolphin}(2012)}]{Dolphin2012}
---. 2012, \apj, 751, 60

\bibitem[{{Dolphin}(2013)}]{Dolphin2013}
---. 2013, \apj, 775, 76

\bibitem[{{Drlica-Wagner} {et~al.}(2015){Drlica-Wagner}, {Bechtol}, {Rykoff},
  {Luque}, {Queiroz}, {Mao}, {Wechsler}, {Simon}, {Santiago}, {Yanny},
  {Balbinot}, {Dodelson}, {Fausti Neto}, {James}, {Li}, {Maia}, {Marshall},
  {Pieres}, {Stringer}, {Walker}, {Abbott}, {Abdalla}, {Allam},
  {Benoit-L{\'e}vy}, {Bernstein}, {Bertin}, {Brooks}, {Buckley-Geer}, {Burke},
  {Carnero Rosell}, {Carrasco Kind}, {Carretero}, {Crocce}, {da Costa},
  {Desai}, {Diehl}, {Dietrich}, {Doel}, {Eifler}, {Evrard}, {Finley},
  {Flaugher}, {Fosalba}, {Frieman}, {Gaztanaga}, {Gerdes}, {Gruen}, {Gruendl},
  {Gutierrez}, {Honscheid}, {Kuehn}, {Kuropatkin}, {Lahav}, {Martini},
  {Miquel}, {Nord}, {Ogando}, {Plazas}, {Reil}, {Roodman}, {Sako}, {Sanchez},
  {Scarpine}, {Schubnell}, {Sevilla-Noarbe}, {Smith}, {Soares-Santos},
  {Sobreira}, {Suchyta}, {Swanson}, {Tarle}, {Tucker}, {Vikram}, {Wester},
  {Zhang}, {Zuntz}, \& {DES Collaboration}}]{Drlica-Wagner2015}
{Drlica-Wagner}, A., {Bechtol}, K., {Rykoff}, E.~S., {et~al.} 2015, \apj, 813,
  109

\bibitem[{{Drlica-Wagner} {et~al.}(2020){Drlica-Wagner}, {Bechtol}, {Mau},
  {McNanna}, {Nadler}, {Pace}, {Li}, {Pieres}, {Rozo}, {Simon}, {Walker},
  {Wechsler}, {Abbott}, {Allam}, {Annis}, {Bertin}, {Brooks}, {Burke},
  {Rosell}, {Carrasco Kind}, {Carretero}, {Costanzi}, {da Costa}, {De Vicente},
  {Desai}, {Diehl}, {Doel}, {Eifler}, {Everett}, {Flaugher}, {Frieman},
  {Garc{\'\i}a-Bellido}, {Gaztanaga}, {Gruen}, {Gruendl}, {Gschwend},
  {Gutierrez}, {Honscheid}, {James}, {Krause}, {Kuehn}, {Kuropatkin}, {Lahav},
  {Maia}, {Marshall}, {Melchior}, {Menanteau}, {Miquel}, {Palmese}, {Plazas},
  {Sanchez}, {Scarpine}, {Schubnell}, {Serrano}, {Sevilla-Noarbe}, {Smith},
  {Suchyta}, {Tarle}, \& {DES Collaboration}}]{DrlicaWagner2020}
{Drlica-Wagner}, A., {Bechtol}, K., {Mau}, S., {et~al.} 2020, \apj, 893, 47

\bibitem[{{Fillingham} {et~al.}(2018){Fillingham}, {Cooper}, {Boylan-Kolchin},
  {Bullock}, {Garrison-Kimmel}, \& {Wheeler}}]{Fillingham2018}
{Fillingham}, S.~P., {Cooper}, M.~C., {Boylan-Kolchin}, M., {et~al.} 2018,
  \mnras, 477, 4491

\bibitem[{{Ford} {et~al.}(1998){Ford}, {Bartko}, {Bely}, {Broadhurst},
  {Burrows}, {Cheng}, {Clampin}, {Crocker}, {Feldman}, {Golimowski}, {Hartig},
  {Illingworth}, {Kimble}, {Lesser}, {Miley}, {Neff}, {Postman}, {Sparks},
  {Tsvetanov}, {White}, {Sullivan}, {Krebs}, {Leviton}, {La Jeunesse},
  {Burmester}, {Fike}, {Johnson}, {Slusher}, {Volmer}, \&
  {Woodruff}}]{Ford1998}
{Ford}, H.~C., {Bartko}, F., {Bely}, P.~Y., {et~al.} 1998, in Society of
  Photo-Optical Instrumentation Engineers (SPIE) Conference Series, Vol. 3356,
  Space Telescopes and Instruments V, ed. P.~Y. {Bely} \& J.~B. {Breckinridge},
  234--248

\bibitem[{{Foreman-Mackey} {et~al.}(2013){Foreman-Mackey}, {Hogg}, {Lang}, \&
  {Goodman}}]{Foreman-Mackey2013}
{Foreman-Mackey}, D., {Hogg}, D.~W., {Lang}, D., \& {Goodman}, J. 2013, \pasp,
  125, 306

\bibitem[{{Garrison-Kimmel} {et~al.}(2019){Garrison-Kimmel}, {Wetzel},
  {Hopkins}, {Sanderson}, {El-Badry}, {Graus}, {Chan}, {Feldmann},
  {Boylan-Kolchin}, {Hayward}, {Bullock}, {Fitts}, {Samuel}, {Wheeler},
  {Kere{\v{s}}}, \& {Faucher-Gigu{\`e}re}}]{Garrison-Kimmel2019}
{Garrison-Kimmel}, S., {Wetzel}, A., {Hopkins}, P.~F., {et~al.} 2019, \mnras,
  489, 4574

\bibitem[{{Geha} {et~al.}(2012){Geha}, {Blanton}, {Yan}, \&
  {Tinker}}]{Geha2012}
{Geha}, M., {Blanton}, M.~R., {Yan}, R., \& {Tinker}, J.~L. 2012, \apj, 757, 85

\bibitem[{{Hack} {et~al.}(2013){Hack}, {Dencheva}, \& {Fruchter}}]{Hack2013}
{Hack}, W.~J., {Dencheva}, N., \& {Fruchter}, A.~S. 2013, in Astronomical
  Society of the Pacific Conference Series, Vol. 475, Astronomical Data
  Analysis Software and Systems XXII, ed. D.~N. {Friedel}, 49

\bibitem[{{Haynes} {et~al.}(2018){Haynes}, {Giovanelli}, {Kent}, {Adams},
  {Balonek}, {Craig}, {Fertig}, {Finn}, {Giovanardi}, {Hallenbeck}, {Hess},
  {Hoffman}, {Huang}, {Jones}, {Koopmann}, {Kornreich}, {Leisman}, {Miller},
  {Moorman}, {O'Connor}, {O'Donoghue}, {Papastergis}, {Troischt}, {Stark}, \&
  {Xiao}}]{Haynes2018}
{Haynes}, M.~P., {Giovanelli}, R., {Kent}, B.~R., {et~al.} 2018, \apj, 861, 49

\bibitem[{{Hidalgo} {et~al.}(2018){Hidalgo}, {Pietrinferni}, {Cassisi},
  {Salaris}, {Mucciarelli}, {Savino}, {Aparicio}, {Silva Aguirre}, \&
  {Verma}}]{Hidalgo2018}
{Hidalgo}, S.~L., {Pietrinferni}, A., {Cassisi}, S., {et~al.} 2018, \apj, 856,
  125

\bibitem[{{Higgs} \& {McConnachie}(2021)}]{Higgs2021}
{Higgs}, C.~R., \& {McConnachie}, A.~W. 2021, \mnras, 506, 2766

\bibitem[{{Kauffmann} {et~al.}(1993){Kauffmann}, {White}, \&
  {Guiderdoni}}]{Kauffmann1993}
{Kauffmann}, G., {White}, S.~D.~M., \& {Guiderdoni}, B. 1993, \mnras, 264, 201

\bibitem[{{Koposov} {et~al.}(2015){Koposov}, {Belokurov}, {Torrealba}, \&
  {Evans}}]{Koposov2015}
{Koposov}, S.~E., {Belokurov}, V., {Torrealba}, G., \& {Evans}, N.~W. 2015,
  \apj, 805, 130

\bibitem[{{Kroupa}(2001)}]{Kroupa2001}
{Kroupa}, P. 2001, \mnras, 322, 231

\bibitem[{{Mapelli} {et~al.}(2009){Mapelli}, {Ripamonti}, {Battaglia},
  {Tolstoy}, {Irwin}, {Moore}, \& {Sigurdsson}}]{Mapelli2009}
{Mapelli}, M., {Ripamonti}, E., {Battaglia}, G., {et~al.} 2009, \mnras, 396,
  1771

\bibitem[{{Mapelli} {et~al.}(2007){Mapelli}, {Ripamonti}, {Tolstoy},
  {Sigurdsson}, {Irwin}, \& {Battaglia}}]{Mapelli2007}
{Mapelli}, M., {Ripamonti}, E., {Tolstoy}, E., {et~al.} 2007, \mnras, 380, 1127

\bibitem[{{Martin} {et~al.}(2008){Martin}, {de Jong}, \& {Rix}}]{Martin2008}
{Martin}, N.~F., {de Jong}, J. T.~A., \& {Rix}, H.-W. 2008, \apj, 684, 1075

\bibitem[{{Martin} {et~al.}(2016){Martin}, {Ibata}, {Lewis}, {McConnachie},
  {Babul}, {Bate}, {Bernard}, {Chapman}, {Collins}, {Conn}, {Crnojevi{\'c}},
  {Fardal}, {Ferguson}, {Irwin}, {Mackey}, {McMonigal}, {Navarro}, \&
  {Rich}}]{Martin2016b}
{Martin}, N.~F., {Ibata}, R.~A., {Lewis}, G.~F., {et~al.} 2016, \apj, 833, 167

\bibitem[{{Mateo} {et~al.}(1995){Mateo}, {Fischer}, \&
  {Krzeminski}}]{Mateo1995}
{Mateo}, M., {Fischer}, P., \& {Krzeminski}, W. 1995, \aj, 110, 2166

\bibitem[{{McConnachie}(2012)}]{McConnachie2012}
{McConnachie}, A.~W. 2012, \aj, 144, 4

\bibitem[{{McConnachie} {et~al.}(2009){McConnachie}, {Irwin}, {Ibata},
  {Dubinski}, {Widrow}, {Martin}, {C{\^o}t{\'e}}, {Dotter}, {Navarro},
  {Ferguson}, {Puzia}, {Lewis}, {Babul}, {Barmby}, {Bienaym{\'e}}, {Chapman},
  {Cockcroft}, {Collins}, {Fardal}, {Harris}, {Huxor}, {Mackey},
  {Pe{\~n}arrubia}, {Rich}, {Richer}, {Siebert}, {Tanvir}, {Valls-Gabaud}, \&
  {Venn}}]{McConnachie2009}
{McConnachie}, A.~W., {Irwin}, M.~J., {Ibata}, R.~A., {et~al.} 2009, \nat, 461,
  66

\bibitem[{{McQuinn} {et~al.}(2011){McQuinn}, {Skillman}, {Dalcanton},
  {Dolphin}, {Holtzman}, {Weisz}, \& {Williams}}]{McQuinn2011}
{McQuinn}, K. B.~W., {Skillman}, E.~D., {Dalcanton}, J.~J., {et~al.} 2011,
  \apj, 740, 48

\bibitem[{{McQuinn} {et~al.}(2013){McQuinn}, {Skillman}, {Berg}, {Cannon},
  {Salzer}, {Adams}, {Dolphin}, {Giovanelli}, {Haynes}, \&
  {Rhode}}]{McQuinn2013}
{McQuinn}, K. B.~W., {Skillman}, E.~D., {Berg}, D., {et~al.} 2013, \aj, 146,
  145

\bibitem[{{McQuinn} {et~al.}(2014){McQuinn}, {Cannon}, {Dolphin}, {Skillman},
  {Salzer}, {Haynes}, {Adams}, {Cave}, {Elson}, {Giovanelli}, {Ott}, \&
  {Saintonge}}]{McQuinn2014}
{McQuinn}, K. B.~W., {Cannon}, J.~M., {Dolphin}, A.~E., {et~al.} 2014, \apj,
  785, 3

\bibitem[{{McQuinn} {et~al.}(2015{\natexlab{a}}){McQuinn}, {Skillman},
  {Dolphin}, {Cannon}, {Salzer}, {Rhode}, {Adams}, {Berg}, {Giovanelli},
  {Girardi}, \& {Haynes}}]{McQuinn2015a}
{McQuinn}, K. B.~W., {Skillman}, E.~D., {Dolphin}, A., {et~al.}
  2015{\natexlab{a}}, \apj, 812, 158

\bibitem[{{McQuinn} {et~al.}(2015{\natexlab{b}}){McQuinn}, {Skillman},
  {Dolphin}, {Cannon}, {Salzer}, {Rhode}, {Adams}, {Berg}, {Giovanelli}, \&
  {Haynes}}]{McQuinn2015b}
---. 2015{\natexlab{b}}, \apjl, 815, L17

\bibitem[{{Miyoshi} \& {Chiba}(2020)}]{Miyoshi2020}
{Miyoshi}, T., \& {Chiba}, M. 2020, \apj, 905, 109

\bibitem[{{Momany} {et~al.}(2007){Momany}, {Held}, {Saviane}, {Zaggia},
  {Rizzi}, \& {Gullieuszik}}]{Momany2007}
{Momany}, Y., {Held}, E.~V., {Saviane}, I., {et~al.} 2007, \aap, 468, 973

\bibitem[{{Monelli} {et~al.}(2012){Monelli}, {Cassisi}, {Mapelli}, {Bernard},
  {Aparicio}, {Skillman}, {Stetson}, {Gallart}, {Hidalgo}, {Mayer}, \&
  {Tolstoy}}]{Monelli2012}
{Monelli}, M., {Cassisi}, S., {Mapelli}, M., {et~al.} 2012, \apj, 744, 157

\bibitem[{{Moore} {et~al.}(1999){Moore}, {Ghigna}, {Governato}, {Lake},
  {Quinn}, {Stadel}, \& {Tozzi}}]{Moore1999}
{Moore}, B., {Ghigna}, S., {Governato}, F., {et~al.} 1999, \apjl, 524, L19

\bibitem[{{Munshi} {et~al.}(2021){Munshi}, {Brooks}, {Applebaum},
  {Christensen}, {Quinn}, \& {Sligh}}]{Munshi2021}
{Munshi}, F., {Brooks}, A.~M., {Applebaum}, E., {et~al.} 2021, \apj, 923, 35

\bibitem[{{Nadler} {et~al.}(2021){Nadler}, {Drlica-Wagner}, {Bechtol}, {Mau},
  {Wechsler}, {Gluscevic}, {Boddy}, {Pace}, {Li}, {McNanna}, {Riley},
  {Garc{\'\i}a-Bellido}, {Mao}, {Green}, {Burke}, {Peter}, {Jain}, {Abbott},
  {Aguena}, {Allam}, {Annis}, {Avila}, {Brooks}, {Carrasco Kind}, {Carretero},
  {Costanzi}, {da Costa}, {De Vicente}, {Desai}, {Diehl}, {Doel}, {Everett},
  {Evrard}, {Flaugher}, {Frieman}, {Gerdes}, {Gruen}, {Gruendl}, {Gschwend},
  {Gutierrez}, {Hinton}, {Honscheid}, {Huterer}, {James}, {Krause}, {Kuehn},
  {Kuropatkin}, {Lahav}, {Maia}, {Marshall}, {Menanteau}, {Miquel}, {Palmese},
  {Paz-Chinch{\'o}n}, {Plazas}, {Romer}, {Sanchez}, {Scarpine}, {Serrano},
  {Sevilla-Noarbe}, {Smith}, {Soares-Santos}, {Suchyta}, {Swanson}, {Tarle},
  {Tucker}, {Walker}, {Wester}, \& {DES Collaboration}}]{Nadler2021}
{Nadler}, E.~O., {Drlica-Wagner}, A., {Bechtol}, K., {et~al.} 2021, \prl, 126,
  091101

\bibitem[{{Okamoto} {et~al.}(2012){Okamoto}, {Arimoto}, {Yamada}, \&
  {Onodera}}]{Okamoto2012}
{Okamoto}, S., {Arimoto}, N., {Yamada}, Y., \& {Onodera}, M. 2012, \apj, 744,
  96

\bibitem[{{Pan} {et~al.}(2023){Pan}, {Simpson}, {Kravtsov}, {G{\'o}mez},
  {Grand}, {Marinacci}, {Pakmor}, {Manwadkar}, \& {Esmerian}}]{Pan2022}
{Pan}, Y., {Simpson}, C.~M., {Kravtsov}, A., {et~al.} 2023, \mnras, 519, 4499

\bibitem[{{Pereira-Wilson} {et~al.}(2023){Pereira-Wilson}, {Navarro},
  {Ben{\'\i}tez-Llambay}, \& {Santos-Santos}}]{Wilson2022}
{Pereira-Wilson}, M., {Navarro}, J.~F., {Ben{\'\i}tez-Llambay}, A., \&
  {Santos-Santos}, I. 2023, \mnras, 519, 1425

\bibitem[{{Rey} {et~al.}(2020){Rey}, {Pontzen}, {Agertz}, {Orkney}, {Read}, \&
  {Rosdahl}}]{Rey2020}
{Rey}, M.~P., {Pontzen}, A., {Agertz}, O., {et~al.} 2020, \mnras, 497, 1508

\bibitem[{{Rodriguez Wimberly} {et~al.}(2019){Rodriguez Wimberly}, {Cooper},
  {Fillingham}, {Boylan-Kolchin}, {Bullock}, \&
  {Garrison-Kimmel}}]{RodriguezWimberly2019}
{Rodriguez Wimberly}, M.~K., {Cooper}, M.~C., {Fillingham}, S.~P., {et~al.}
  2019, \mnras, 483, 4031

\bibitem[{{Sand} {et~al.}(2010){Sand}, {Seth}, {Olszewski}, {Willman},
  {Zaritsky}, \& {Kallivayalil}}]{Sand2010}
{Sand}, D.~J., {Seth}, A., {Olszewski}, E.~W., {et~al.} 2010, \apj, 718, 530

\bibitem[{{Sand} {et~al.}(2022){Sand}, {Mutlu-Pakdil}, {Jones}, {Karunakaran},
  {Wang}, {Yang}, {Chiti}, {Bennet}, {Crnojevi{\'c}}, \& {Spekkens}}]{Sand2022}
{Sand}, D.~J., {Mutlu-Pakdil}, B., {Jones}, M.~G., {et~al.} 2022, \apjl, 935,
  L17

\bibitem[{{Santana} {et~al.}(2013){Santana}, {Mu{\~n}oz}, {Geha},
  {C{\^o}t{\'e}}, {Stetson}, {Simon}, \& {Djorgovski}}]{Santana2013}
{Santana}, F.~A., {Mu{\~n}oz}, R.~R., {Geha}, M., {et~al.} 2013, \apj, 774, 106

\bibitem[{{Savino} {et~al.}(2022){Savino}, {Weisz}, {Skillman}, {Dolphin},
  {Kallivayalil}, {Wetzel}, {Anderson}, {Besla}, {Boylan-Kolchin}, {Bullock},
  {Cole}, {Collins}, {Cooper}, {Deason}, {Dotter}, {Fardal}, {Ferguson},
  {Fritz}, {Geha}, {Gilbert}, {Guhathakurta}, {Ibata}, {Irwin}, {Jeon},
  {Kirby}, {Lewis}, {Mackey}, {Majewski}, {Martin}, {McConnachie}, {Patel},
  {Rich}, {Simon}, {Sohn}, {Tollerud}, \& {van der Marel}}]{Savino2022}
{Savino}, A., {Weisz}, D.~R., {Skillman}, E.~D., {et~al.} 2022, \apj, 938, 101

\bibitem[{{Schlafly} \& {Finkbeiner}(2011)}]{Schlafly2011}
{Schlafly}, E.~F., \& {Finkbeiner}, D.~P. 2011, \apj, 737, 103

\bibitem[{{Schlegel} {et~al.}(1998){Schlegel}, {Finkbeiner}, \&
  {Davis}}]{Schlegel1998}
{Schlegel}, D.~J., {Finkbeiner}, D.~P., \& {Davis}, M. 1998, \apj, 500, 525

\bibitem[{{Simon}(2019)}]{Simon2019}
{Simon}, J.~D. 2019, \araa, 57, 375

\bibitem[{{Sirianni} {et~al.}(2005){Sirianni}, {Jee}, {Ben{\'\i}tez},
  {Blakeslee}, {Martel}, {Meurer}, {Clampin}, {De Marchi}, {Ford}, {Gilliland},
  {Hartig}, {Illingworth}, {Mack}, \& {McCann}}]{Sirianni2005}
{Sirianni}, M., {Jee}, M.~J., {Ben{\'\i}tez}, N., {et~al.} 2005, \pasp, 117,
  1049

\bibitem[{{Skillman} {et~al.}(2017){Skillman}, {Monelli}, {Weisz}, {Hidalgo},
  {Aparicio}, {Bernard}, {Boylan-Kolchin}, {Cassisi}, {Cole}, {Dolphin},
  {Ferguson}, {Gallart}, {Irwin}, {Martin}, {Mart{\'\i}nez-V{\'a}zquez},
  {Mayer}, {McConnachie}, {McQuinn}, {Navarro}, \& {Stetson}}]{Skillman2017}
{Skillman}, E.~D., {Monelli}, M., {Weisz}, D.~R., {et~al.} 2017, \apj, 837, 102

\bibitem[{{Somerville}(2002)}]{Somerville2002}
{Somerville}, R.~S. 2002, \apjl, 572, L23

\bibitem[{{Telford} {et~al.}(2020){Telford}, {Dalcanton}, {Williams}, {Bell},
  {Dolphin}, {Durbin}, \& {Choi}}]{Telford2020}
{Telford}, O.~G., {Dalcanton}, J.~J., {Williams}, B.~F., {et~al.} 2020, \apj,
  891, 32

\bibitem[{{Teyssier} {et~al.}(2012){Teyssier}, {Johnston}, \&
  {Kuhlen}}]{Teyssier2012}
{Teyssier}, M., {Johnston}, K.~V., \& {Kuhlen}, M. 2012, \mnras, 426, 1808

\bibitem[{{The Astropy Collaboration}(2018)}]{astropy:2018}
{The Astropy Collaboration}. 2018, {astropy v3.1: a core python package for
  astronomy}, Zenodo, doi:10.5281/zenodo.4080996

\bibitem[{{Vincenzo} {et~al.}(2016){Vincenzo}, {Matteucci}, {Belfiore}, \&
  {Maiolino}}]{Vincenzo2016}
{Vincenzo}, F., {Matteucci}, F., {Belfiore}, F., \& {Maiolino}, R. 2016,
  \mnras, 455, 4183

\bibitem[{{Weisz} {et~al.}(2019){Weisz}, {Martin}, {Dolphin}, {Albers},
  {Collins}, {Ferguson}, {Lewis}, {Mackey}, {McConnachie}, {Rich}, \&
  {Skillman}}]{Weisz2019}
{Weisz}, D.~R., {Martin}, N.~F., {Dolphin}, A.~E., {et~al.} 2019, \apjl, 885,
  L8

\bibitem[{{Wetzel} {et~al.}(2015){Wetzel}, {Tollerud}, \& {Weisz}}]{Wetzel2015}
{Wetzel}, A.~R., {Tollerud}, E.~J., \& {Weisz}, D.~R. 2015, \apjl, 808, L27

\bibitem[{{Williams} {et~al.}(2014){Williams}, {Lang}, {Dalcanton}, {Dolphin},
  {Weisz}, {Bell}, {Bianchi}, {Byler}, {Gilbert}, {Girardi}, {Gordon},
  {Gregersen}, {Johnson}, {Kalirai}, {Lauer}, {Monachesi}, {Rosenfield},
  {Seth}, \& {Skillman}}]{Williams2014}
{Williams}, B.~F., {Lang}, D., {Dalcanton}, J.~J., {et~al.} 2014, \apjs, 215, 9

\bibitem[{{Williams} {et~al.}(2021){Williams}, {Durbin}, {Dalcanton}, {Lang},
  {Girardi}, {Smercina}, {Dolphin}, {Weisz}, {Choi}, {Bell}, {Rosolowsky},
  {Skillman}, {Koch}, {Lindberg}, {Hagen}, {Gordon}, {Seth}, {Gilbert},
  {Guhathakurta}, {Lauer}, \& {Bianchi}}]{Williams2021}
{Williams}, B.~F., {Durbin}, M.~J., {Dalcanton}, J.~J., {et~al.} 2021, \apjs,
  253, 53

\bibitem[{{York} {et~al.}(2000){York}, {Adelman}, {Anderson}, {Anderson},
  {Annis}, {Bahcall}, {Bakken}, {Barkhouser}, {Bastian}, {Berman}, {Boroski},
  {Bracker}, {Briegel}, {Briggs}, {Brinkmann}, {Brunner}, {Burles}, {Carey},
  {Carr}, {Castander}, {Chen}, {Colestock}, {Connolly}, {Crocker}, {Csabai},
  {Czarapata}, {Davis}, {Doi}, {Dombeck}, {Eisenstein}, {Ellman}, {Elms},
  {Evans}, {Fan}, {Federwitz}, {Fiscelli}, {Friedman}, {Frieman}, {Fukugita},
  {Gillespie}, {Gunn}, {Gurbani}, {de Haas}, {Haldeman}, {Harris}, {Hayes},
  {Heckman}, {Hennessy}, {Hindsley}, {Holm}, {Holmgren}, {Huang}, {Hull},
  {Husby}, {Ichikawa}, {Ichikawa}, {Ivezi{\'c}}, {Kent}, {Kim}, {Kinney},
  {Klaene}, {Kleinman}, {Kleinman}, {Knapp}, {Korienek}, {Kron}, {Kunszt},
  {Lamb}, {Lee}, {Leger}, {Limmongkol}, {Lindenmeyer}, {Long}, {Loomis},
  {Loveday}, {Lucinio}, {Lupton}, {MacKinnon}, {Mannery}, {Mantsch}, {Margon},
  {McGehee}, {McKay}, {Meiksin}, {Merelli}, {Monet}, {Munn}, {Narayanan},
  {Nash}, {Neilsen}, {Neswold}, {Newberg}, {Nichol}, {Nicinski}, {Nonino},
  {Okada}, {Okamura}, {Ostriker}, {Owen}, {Pauls}, {Peoples}, {Peterson},
  {Petravick}, {Pier}, {Pope}, {Pordes}, {Prosapio}, {Rechenmacher}, {Quinn},
  {Richards}, {Richmond}, {Rivetta}, {Rockosi}, {Ruthmansdorfer}, {Sandford},
  {Schlegel}, {Schneider}, {Sekiguchi}, {Sergey}, {Shimasaku}, {Siegmund},
  {Smee}, {Smith}, {Snedden}, {Stone}, {Stoughton}, {Strauss}, {Stubbs},
  {SubbaRao}, {Szalay}, {Szapudi}, {Szokoly}, {Thakar}, {Tremonti}, {Tucker},
  {Uomoto}, {Vanden Berk}, {Vogeley}, {Waddell}, {Wang}, {Watanabe},
  {Weinberg}, {Yanny}, {Yasuda}, \& {SDSS Collaboration}}]{York2000}
{York}, D.~G., {Adelman}, J., {Anderson}, John~E., J., {et~al.} 2000, \aj, 120,
  1579

\end{thebibliography}

\end{document}